# THE NEXT-TO-LEADING ORDER OF THE DIFFERENTIAL CROSS-SECTION OF THE SUBPROCESS OF ANNIHILATION OF QUARK-ANTIQUARK PAIR $q\bar{q} \to g\gamma$ OF PROMPT PHOTON PRODUCTION IN PROTON-PROTON COLLISIONS AT $\sqrt{s}$=10 GeV NICA ENERGIES

**M.R.Alizada[1], A.I.Ahmadov[1,2]**

[1]Department of Theoretical Physics, Baku State University
Z.Khalilov 33, Az-1148 Baku, Azerbaijan, mohsunalizade@gmail.com
[2]Institute for Physical Problems, Baku State University,
Z.Khalilov 33, Az-1148 Baku, Azerbaijan

## ABSTRACT

Next-to-the leading order of the differential cross-section of the subprocess of annihilation of quark-antiquark pair $q\bar{q} \to g\gamma$ of prompt photon production in nonpolarized and longitudinally polarized proton-proton collisions at NICA energies has been investigated.

Shown, that contribution of the next-to-leading order to the differential cross-section of annihilation of quark-antiquark pair $q\bar{q} \to g\gamma$ is significant at the high energies of colliding protons and consist around 15% of leading order calculation.

The influence of longitudinal polarization of colliding protons on the contributions of the next-to-leading order is more significant than, on the contributions of the leading order and it is significant at high energies of colliding protons.

PYTHIA 8.316 + POWHEG BOX Monte Carlo (MC) event generator has been used for numerically simulation of prompt photon production in annihilation of quark-antiquark pair in NLO approximation for determination of the dependences of cross section on the sum of energies of colliding protons $\sqrt{s}$, transverse momentum $p_T$, cosine of scattering angle $Cos(\theta)$, rapidity y of photon and $x_T$. The NLO+PS simulation accurately reproduces the fundamental signatures of pQCD.

## INTRODUCTION

The production of prompt photons in proton-proton collisions is an important process for studying the dynamic structure of proton at high energies. Being neutral in color, they have a large mean free path in both dense hadronic matter and quark-gluon plasma (QGP). This allows them to escape from the interaction region unchanged. A photon produced in a proton-proton collision in the energy range from 1.5 to several GeV also carries information about the formation of the QGP, the distribution of partons in nucleons as they are formed, hard scattering of incoming partons such as Compton scattering of quark-gluon $qg \rightarrow g\gamma$, annihilation of quark-antiquark pair $q\bar{q} \rightarrow g\gamma$, and bremsstrahlung $qq \rightarrow qq\gamma$ subjected to hard scattering [1-3].

At early, theoretical and experimental investigations of the prompt photon production in the proton-proton collisions was been carried out at the LHC and American Tevatron energies. It is shown that the subprocesses of Compton scattering of a quark-gluon $qg \rightarrow g\gamma$ and the annihilation of a quark-antiquark pair $q\bar{q} \rightarrow g\gamma$ are the basic mechanisms for the production of prompt photons. The mixed QCD Compton scattering subprocess is determined to be the dominant process of prompt photon production in proton-proton collisions [4,5].

It is known that, the accuracy of theoretical calculations of the prompt photon production depends on the order of the perturbative expansion, which is usually expressed in terms of the strong coupling constant $\alpha_s$. The leading order (LO) approximation, which corresponds to the lowest power of strength coupling $\alpha_s$, is often insufficient to describe the experimental data, especially at large transverse momenta of photons $p_T$. Therefore, it is necessary to use higher-order corrections, such as next-to-leading order (NLO), next-to-next-to-leading order (NNLO), which take into account the effects of one-loop virtual diagrams and real production of additional particles [6-12].

In the [6] are presented results of theoretical studies of the isolated production of prompt photons at the DESY HERA collider in the NLO quantum chromodynamics (QCD). The authors used analytical methods to calculate the differential cross-sections of various subprocesses and applied isolating cross-section, corresponding to the experimental conditions. It was shown that the differential cross-section is sensitive to the parton distribution function (PDF) of both the photon and the quark, but the gluon distribution is difficult to measure.

The authors of [7] presented a theoretical calculation of the photon production rate resulting from annihilation of quark-antiquark, specifically considering the interaction of a charm quark ($c$) and an anti-strange quark ($\bar{s}$). The aim was to study the influence of various parameters on this rate, including the QCD strength coupling $\alpha_s$, the critical temperature ($T_c$) of deconfinement, the photon energy ($E_y$), and the thermal energy (temperature $T$) of the QGP system. Calculations were performed for two different critical temperatures ($T_c$=113 MeV and $T_c$=147 MeV) and over a photon energy range from 1.5 GeV/c to 5 GeV/c. An increased photon rate with decreased QCD strength coupling $\alpha_s$ of quark interaction due to increasing critical temperature was predicted. It was noted, that the highest photon production rate is observed at photon energies $E_y \leq 2$ GeV. The influence of critical temperature $T_c$, thermal energy $T$, photon energy $E_y$, and QCD strength coupling constant $\alpha_s$ on the photon production rate was investigated in detail.

The authors [8] computed for the first time the so-called "complete" NLO corrections to top-quark pair production in association with one and two isolated photons ($tt\gamma$ and $tt\gamma\gamma$) considering the di-lepton decay channel of the top quarks. A key feature is the consistent inclusion of higher-order QCD and electroweak (EW) effects, as well as photon bremsstrahlung, at all stages of the process: in the production of the top-quark pair and in their decays.

In this article are solved following tasks:

1. To present the first complete NLO calculations for $pp \to tt\gamma$ and $pp \to tt\gamma\gamma$ in the di-lepton top-quark decay channel;
2. To model unstable top quarks and W bosons using the Narrow Width Approximation (NWA);
3. To present results at the integrated and differential fiducial cross-section level for the LHC Run II center-of-mass energy ($\sqrt{s}$=13 TeV);
4. To investigate the scale choice in photonic observables;
5. To discuss in detail the individual size of each subleading contribution and scrutinize the origin of the main subleading corrections.

The work provides more precise theoretical predictions for important processes at the LHC, which serve as signals for studying top-quark properties and as backgrounds for new physics searches and Higgs boson studies.

The author of the dissertation [9] presents a calculation of prompt photon production in hadron collisions at NLO in perturbative QCD. The work focuses on predictions at low photon transverse momenta $p_T$ using the JETPHOX program and on matching these calculations to parton shower generators using the POWHEG BOX framework.

1. To perform NLO QCD calculations for prompt photon production;
2. To investigate the theoretical uncertainties of these calculations;
3. To verify and confirm the compatibility of ALICE experiment data on direct photons with the existence of an excess of photons, which can be interpreted as a thermal photon signal originating from the strongly interacting medium produced in heavy-ion collisions.

To match NLO calculations with parton shower generators (using POWHEG BOX) and to show that this qualitatively improves the description of PHENIX experiment data compared to fixed-order calculations:

1. NLO QCD calculations for prompt photon production, including the fragmentation contribution, are essential for describing data at low transverse momenta $p_T$;
2. ALICE data on direct photons in $Pb - Pb$ collisions are consistent with the existence of a thermal photon signal, indicating the formation of a hot and dense medium;
3. Matching NLO calculations with parton showers using POWHEG BOX allows for more realistic predictions for exclusive observables and qualitatively improves the description of PHENIX data compared to fixed-order calculations.

This work represents an important step in developing precise theoretical tools for analyzing processes with prompt photons in both proton-proton and heavy-ion collisions.

In the [10], the authors investigate the production of direct photon pairs in hadron collisions in the energy range from fixed target to LHC energies. The DIPHOX code allows all contributions to photon pair production to be taken into account with NLO accuracy, including fragmentation contributions. NLO calculations agree well with experimental data in the kinematic region where they are applicable. For the LHC, NLO predictions provide significant corrections to the leading order, but uncertainties related to the choice of scale remain significant. Some infrared-sensitive observables require further theoretical study.

The authors [11] computed the NNLO QCD corrections to the cross sections for isolated single photon and photon-plus-jet production at hadron colliders. A key feature of this work is the inclusion of two types of contributions that arise with realistic photon isolation criteria:

1. Direct photon production;
2. Parton-to-photon fragmentation.

Previous NNLO calculations used idealized (dynamical or hybrid) isolation criteria, which excluded the fragmentation contribution. This work presents the first NNLO calculations using the standard fixed-cone isolation procedure, which is applied in experiments. This allows for more accurate and direct comparisons of theoretical predictions with LHC experimental data.

The objectives were:

1. To provide NNLO QCD predictions for photon and photon+jet production with realistic isolation;
2. To compare these predictions with data from ATLAS, ALICE, and CMS experiments;
3. To quantify the impact of different photon isolation prescriptions (including no isolation) on the cross-sections.

To discuss the possibility of measuring photon fragmentation functions (FF) at hadron colliders using a new observable,

1. The authors presented the first NNLO QCD calculations for single photon and photon+jet production with realistic fixed-cone isolation, including fragmentation contributions;
2. The new results allow for more accurate comparisons with LHC experimental data, eliminating the systematic uncertainty arising from the mismatch of isolation criteria in theory and experiment;
3. The impact of NNLO corrections on various observables and improved agreement with data were demonstrated;
4. The influence of different isolation criteria was investigated, showing that democratic clustering can closely reproduce fixed-cone isolation results.

The observable was proposed and studied, which could be used to measure photon FF at the LHC, particularly in scenarios without photon isolation.

Thus, may be say that, NLO and NNLO calculations are more complicated than LO calculations, because they include ultraviolet and infrared divergences, which need to be regularized and renormalized. NLO results depend on the choice of renormalization and factorization scales, which introduce theoretical uncertainties, which need to be estimated [12,13].

Due to the high energy of the colliding nucleons, many different elementary particles are produced in addition to prompt photons, and consequently the production of additional elementary particles complicates the accuracy of determining the effective cross-section for the production of prompt photons. From this point of view, research into the production of prompt photons in proton-proton collisions at the Nuclotron-based Ion Collider fAcility, i.e. NICA, energies has its advantages. The energy 10 GeV NICA does not allow the production of many elementary particles, which makes it difficult to accurately determine the differential cross-section of prompt photons production, and will also help determine the temperature of the phase transition to quark-gluon plasma and hadronization [14,15].

At early, we investigated differential cross-section of the subprocesses Compton scattering quark-gluon $qg \rightarrow q\gamma$, annihilation of quark-antiquark pairs $q\bar{q} \rightarrow g\gamma$ and bremsstrahlung $qq \rightarrow qq\gamma$ prompt photon production in proton-proton collision at NICA energies in LO approximation [16-18]. Differential cross-section of prompt photons production consists of more than 50% of the differential cross section of the subprocess of Compton scattering of quark-gluon $qg \rightarrow q\gamma$ and about 48% of the differential cross section of annihilation of quark-antiquark pairs $q\bar{q} \rightarrow g\gamma$ and less than 0.03% of the differential cross section of bremsstrahlung $qq \rightarrow qq\gamma$. Despite the low collision energies, Compton scattering of quark-gluon $qg \rightarrow g\gamma$ remains dominant. We also have shown that the prompt photons produced in pure electrodynamics Compton scattering constitute 10% of the prompt photons produced in mixed Compton scattering [16-18]. We investigated prompt photon production in Compton quark-gluon scattering $qg \rightarrow q\gamma$ in the NLO approximation at collision of nonpolarized and the longitudinally polarized protons, and compares with calculations in LO [17].

In the presented article, we continued the study of prompt photon production in annihilation of quark-antiquark pair $q\bar{q} \rightarrow g\gamma$ in the NLO approximation at collision of nonpolarized and the longitudinally polarized protons, and compares with calculations in LO and simulation in PYTHIA 8.316.

## II. DIFFERENTIAL CROSS-SECTION OF ONE-LOOP CORRECTIONS TO THE SUBPROCESS OF ANNIHILATION OF QUARK-ANTIQUARK PAIR $q\bar{q} \rightarrow g\gamma$

### 2.1 Differential cross section of subprocess annihilation of quark-antiquark pair $q\bar{q} \rightarrow g\gamma$ at collisions nonpolarized protons

We used FeynCalc for generation of Feynman diagrams of the subprocess of annihilation of quark-antiquark pairs $q\bar{q} \rightarrow g\gamma$ in NLO. In this case 773 one-loop diagrams were generated. Among the diagrams, the ones with heavy particles (H,G,Z,W,t,c,b,μ,τ,e), which are not registered at NICA SPD, the ones with triangular loops of fermions and vector bosons, which vanish by the Furry theorem [18], the ones with "seagulls" and the ones with loops on external legs, which are excluded in the renormalization scheme on the mass shell, as well as the ones with photon-gluon transitions, which do not conserve color, were not taken into account. As a result, we obtained the following 13 diagrams, which contribute to the subprocess annihilation of quark-antiquark pair of prompt photons production in NLO at NICA energies. The Feynman diagrams and matrix elements of subprocess are presented in Appendix.

NLO calculation has been carried out using dimensional regularization and the Minimal Supersymmetric Standard Model (MSSM) scheme [19].

The square of the modulus of the matrix element for the one-loop correction subprocess $q\bar{q} \to g\gamma$, with applied dimensional regularization and averaged over the spin of the initial particles, has the form:

$$|\bar{M}|^2 = \left[\frac{0.00858899\, e^6 e_q^2 g_s^2\, \hat{t}}{\hat{u}} + \frac{e^4 e_q^2 g_s^4}{\hat{t}\,\hat{u}}(-0.167486\,\hat{s}^2 - 0.167486\,\hat{s}\hat{t} + 0.167486\,\hat{t}^2 - 0.167486\,\hat{s}\hat{u}) + \right.$$

$$\left. + \frac{e^2\, e_q^2 g_s^6}{\hat{t}\,\hat{u}}(-1.63299\,\hat{s}^2 - 1.63299\,\hat{s}\hat{t} + 0.816496\,\hat{t}^2 - 1.63299\,\hat{s}\hat{u} + 0.816496\,\hat{u}^2)\right] \times$$

$$\times \left\{1 + \frac{\alpha_s(\mu_R)}{2\pi}\left[C_F \ln\left(\frac{-\hat{t}}{\hat{s}}\right) + C_A \ln\left(\frac{-\hat{u}}{\hat{s}}\right) + \beta_0 \ln\left(\frac{\mu_R^2}{\hat{s}}\right) + C_{\text{virt}}\right]\right\}$$

$$C_F = \frac{N_c^2 - 1}{2N_c}, \qquad C_A = N_c, \qquad \beta_0 = \frac{11}{3}C_A - \frac{4}{3}T_R n_f, \quad T_R = \frac{1}{2}.$$

where $C_F ln(\frac{-\hat{t}}{\hat{s}})$, $C_A \ln(\frac{-\hat{u}}{\hat{s}})$ -- quark color logarithm, gluon color logarithm, correspondingly, $\beta_0 \ln(\frac{\mu_R^2}{\hat{s}})$ - logarithm of the coupling constant renormalization, $\mu_R^2$ - square of the renormalization scale, $C_{virt}$ - finite virtual constant, $g_s$ - is the strong coupling constant at the factorization scale, which we choose equal to $\hat{s}$, $e_q$ - is the electric charge of the quark.

The differential cross-section of the subprocess in parton level was defined as in [20]:

$$\frac{d\hat{\sigma}_{NLO}(q\bar{q} \to g\gamma)}{d\hat{t}} = \left[\frac{0.00858899\, e^6\, e_q^2 g_s^2 t}{16\pi\hat{s}^2\,\hat{u}} + \right.$$

$$+ \frac{e^4\, e e_q^2 g_s^4}{16\pi\hat{s}^2\,\hat{t}\,\hat{u}}(-0.167486\,\hat{s}^2 - 0.167486\,\hat{s}\hat{t} + 0.167486\,\hat{t}^2 - 0.167486\,\hat{s}\hat{u}) +$$

$$\left. + \frac{e^2 e_q^2 g_s^6}{16\pi\hat{s}^2\,\hat{t}\,\hat{u}}(-1.63299\,\hat{s}^2 - 1.63299\,\hat{s}\hat{t} + 0.816496\,\hat{t}^2 - 1.63299\,\hat{s}\hat{u} + 0.816496\,\hat{u}^2)\right]$$

$$\times \left\{1 + \frac{\alpha_s(\mu_R)}{2\pi}\left[C_F \ln\left(\frac{-\hat{t}}{\hat{s}}\right) + C_A \ln\left(\frac{-\hat{u}}{\hat{s}}\right) + \beta_0 \ln\left(\frac{\mu_R^2}{\hat{s}}\right) + C_{\text{virt}}\right]\right\}$$

To calculate the differential cross-section at the hadron level, we used the parton distribution functions (PDF) in NLO level: $G_{q_1/h_1}(x_1)$ and $G_{q_2/h_2}(x_2)$ in accordance with [21] at the factorization scale $\hat{s}$:

$$d\sigma = \int d\hat{\sigma} \cdot G_{q_1/h_1}(x_1) G_{q_2/h_2}(x_2) dx_1 dx_2.$$

LHAPDF 6.5.5 served as our framework for coherent analysis of PDFs under varying scale conditions. The software provided integration with current-generation PDFs libraries that naturally accommodate scale-evolution mechanisms [22]. Through this approach, we successfully investigated how scale transformations affect both the overall normalization and geometric properties of differential measurements.

In particular, for the differential distributions by rapidity $y$ and square of transverse momentum $p_T$, we obtain the following expressions:

$$\frac{d\sigma}{dy} = \int (-t) G_{q_i/h_1}(x_1) G_{q_i/h_2}(x_2) dx_1 dx_2 \frac{d\hat{\sigma}}{d\hat{t}} +$$

$$+ \int (-t) G_{q_i/h_2}(x_1) G_{q_i/h_1}(x_2) dx_1 dx_2 \frac{d\hat{\sigma}}{d\hat{t}} =$$

$$= \int (-t) \left[ G_{q_i/h_1}(x_1) G_{q_i/h_2}(x_2) + G_{q_i/h_2}(x_1) G_{q_i/h_1}(x_2) \right] dx_1 dx_2 \frac{d\hat{\sigma}}{d\hat{t}} =$$

$$= \int (-t) P_{q\bar{q}} dx_1 dx_2 \frac{d\hat{\sigma}}{d\hat{t}}$$

where

$$P_{q\bar{q}} = \sum_{i=1}^{n} e_{q_i}^2 \left[ G_{q_i/h_1}(x_1) G_{q_i/h_2}(x_2) + G_{q_i/h_2}(x_1) G_{q_i/h_1}(x_2) \right].$$

and

$$\frac{d\sigma}{dp_T^2} = \int \left( \frac{-t}{2p_T^2} \right) G_{q_1/h_1}(x_1) G_{q_2/h_2}(x_2) dx_1 dx_2 \frac{d\hat{\sigma}}{d\hat{t}}$$

In a similar way we obtain $d\sigma/dx_T^2$ и $d\sigma/dCos(\theta)$:

$$\frac{d\sigma}{dx_T^2} = \int dx_1 dx_2 \, G_{q_1/h_1}(x_1) \, G_{q_2/h_2}(x_2) \, \frac{d\hat{\sigma}}{d\hat{t}} \, \frac{\hat{s}}{4|Cos(\theta)|},$$

$$\frac{d\sigma}{d\mathrm{Cos}\theta} = \int dx_1 dx_2 \, G_{q_1/h_1}(x_1) \, G_{q_2/h_2}(x_2) \, \frac{d\hat{\sigma}}{d\hat{t}} \, \frac{\hat{s}}{2}$$

2.2 Differential cross section of subprocess annihilation of quark-antiquark pair $q\bar{q} \to g\gamma$ at collisions longitidunally polarized protons

The longitudinal polarization of colliding quarks and antiquarks is taken into account as in [23]:

$$U(p_1)\bar{U}(p_1) = \frac{1}{2}(1 - \lambda_1 \gamma_5)(\hat{p}_1 + m_1),$$

$$v(p_2)\bar{v}(p_2) = \frac{1}{2}(1 + \lambda_2\gamma_5)(\hat{p}_2 + m_2),$$

where $\lambda_1$ и $\lambda_2$ helicity of colliing protons, $m_1$, $m_2$ mass of quarks and antiquark, $p_1$, $p_2$ - transverse momentum of quarks and antiquark, correspondingly.

To calculation of the differential cross section of processes with longitudinally polarized colliding protons a polarized PDF was used as in [24]:

$$\Delta u_v(x, Q_0^2) = x^{-0.441} \cdot (1 - x)^{3.96} \cdot (0.928 + 0.149 \cdot x^{0.5} - 1.141 \cdot x + 11.612 \cdot x^{1.5}),$$

$$\Delta d_v(x, Q_0^2) = x^{-0.665} \cdot (1 - x)^{4.46} \cdot (-0.038 - 0.43 \cdot x^{0.5} - 5.260 \cdot x + 8.443 \cdot x^{1.5}),$$

$$\Delta G(x, Q_0^2) = x^{-1.17} \cdot (1 - x)^{5.33} \cdot (0.03 - 1.71 \cdot x^{0.5} + 3.01 \cdot x + 43.5 \cdot x^{1.5}).$$

It is known that the fraction of photon-type partons in a proton can be comparable to the fraction of sea quarks at large values $x$ [25,26]. The distribution of polarized photons in the proton was taken into account as in [26]:

$$G_{\gamma/p}(x, Q_0) = \frac{\alpha_\gamma}{2\pi}\left(A_u e_u^2 \tilde{P}_{\gamma q} \otimes u(x) + A_d e_d^2 \tilde{P}_{\gamma q} \otimes d(x)\right),$$

where: $A_i = \ln\left(\frac{Q_0^2}{m_i^2}\right)$, $\tilde{P}_{\gamma q} = \frac{1+(1-z)^2}{z}$ - splitting function, $Q_0$ - factorization scale.

The matrix elements and Mandelstam invariants of the process of production of prompt photons in the subprocess of annihilation of the quark-antiquark pair $q\bar{q} \to g\gamma$ studied in the NLO will remain unchanged, as for the case without polarization.

For the square of the matrix element averaged over the spin of the initial particles in the case of one-loop correction of the subprocess $q\bar{q} \to g\gamma$, applying dimensional regularization, we obtain:

$$|\bar{M}|^2 = -\frac{\alpha_s e^2 e_q^2 (1-\lambda_1\lambda_2)(\alpha_s^2(0.0004\hat{s}^2 + 0.0008\hat{s}\hat{t} + 0.0008\hat{t}^2) + 0.001\alpha_s \hat{s}^2\hat{t}^2 + 0.0006\hat{s}^2\hat{t}^2)}{\hat{t}(\hat{s}+\hat{t})}..$$

For the differential cross section of the subprocess $q\bar{q} \to g\gamma$ at the parton level, taking into account the longitudinal polarization of protons, it was obtained as:

$$\frac{d\hat{\sigma}_{NLO}(q\bar{q}\to g\gamma)}{d\hat{t}} = \frac{1}{16\pi\hat{s}^2}\,|\bar{M}|^2.$$

To calculate the differential cross-section at the hadron level, it is necessary to take into account the polarized PDFs for the NLO in the calculations [24].

Double spin asymmetry of subprocesses $q\bar{q} \to g\gamma$ has been calculated as in [25]:

$$A_{LL} = \frac{\sigma^{\uparrow\uparrow} - \sigma^{\uparrow\downarrow}}{\sigma^{\uparrow\uparrow} + \sigma^{\uparrow\downarrow}},$$

where $\sigma^{\uparrow\uparrow}$ и $\sigma^{\uparrow\downarrow}$ - cross sections of processes calculated with, respectively, directed and oppositely directed polarizations of colliding protons.

Double spin asymmetry of subprocesses $q\bar{q} \to g\gamma$ in parton level is obtained:

$$\hat{A}_{LL} = -\lambda_1 \lambda_2.$$

As can be seen, $A_{LL}$ is determined by the product of the helicities of the initial particles.

### 2.3 Simulation of subprocess annihilation of quark-antiquark pair $q\bar{q} \to g\gamma$ in PYHIA 8.316 + POWHEG-BOX

We used sharing the PYTHIA 8.316 + POWHEG-BOX Monte Carlo (MC) event generator for numerically simulate prompt photon production in the annihilation of quark-antiquark pair subprocess for determination of the dependences of cross section on the sum of energies of colliding protons $\sqrt{s}$, transverse momentum $p_T$, cosine of scattering angle $Cos(\theta)$, rapidity y of photon and $x_T$.

MC simulation details in PYTHIA 8.316 + POWHEG-BOX are followings:

To simulate prompt direct photon production in proton-proton $(p + p)$ collisions under the kinematic conditions of the NICA experiment, a two-step framework combining NLO matrix element calculations with a Parton Shower (PS) and hadronization algorithm was employed. The NLO hard scattering events were generated using the POWHEG-BOX-V2 framework, specifically utilizing the gamj process directory. Subsequent parton showering, hadronization, and event analysis were performed using PYTHIA 8.316.

The hard process was calculated at NLO accuracy for $p + p$ collisions at a center-of-mass energy of $\sqrt{s} = 10$ GeV ($E_{\text{beam1}} = E_{\text{beam2}} = 5.0$ GeV), corresponding to the NICA energy regime. A total sample of $N_{\text{evts}} = 10^4$ events was generated.

The initial-state hadron structure was described using the CT14 NLO PDF set (CT14nlo, LHAPDF ID: 13000) for both incoming protons (ih1=1, ih2=1).

To regulate infrared and collinear divergences in the matrix elements, a transverse momentum cut of $p_{T,\text{min}}^{\text{Born}} = 0.5$ GeV$/c$ (bornktmin) was enforced on the Born-level photon/parton. The minimum scale squared for parton shower initiation was configured as $p_{T,\text{min}}^2 = 0.25$ GeV$^2/c^2$ (ptsqmin).

The renormalization and factorization scales were set to the default central hard scale $\mu_0$ of the process (renscfact = 1d0 and facscfact = 1d0, respectively).

The numerical integration over phase space was carried out using 50,000 calls over 5 iterations for both Phase 1 (ncall1 = 50000, itmx1=5) and Phase 2 (ncall2 = 50000, itmx2=5). Integration folding parameters along the variables $\xi$, $y$, and $\phi$ were all set to unity (foldcsi=1, foldy=1, foldphi=1).

The generated hard events were stored in standard Les Houches Event (LHE) format (pwgevents.lhe) for further processing.

The Les Houches event file generated by POWHEG-BOX was imported into PYTHIA 8.316 using the LHE interface (Beams:frameType=4, Beams:LHEF = pwgevents.lhe).

To prevent double-counting of radiation between the NLO matrix element calculations performed in POWHEG-BOX and the parton shower generated by PYTHIA 8.316, the dedicated POWHEG matching/veto scheme was explicitly activated:

1. POWHEG:veto=1: Enables the veto on emissions harder than the POWHEG scale.
2. POWHEG:pTdef=1: Defines the transverse momentum definition for emissions matching the POWHEG internal convention.
3. POWHEG:emitted=0, POWHEG:pTemt=0, POWHEG:pThard=0: Standard matching options specifying how emission hardness is tracked.
4. POWHEG:vetoCount=100: Sets the maximum number of radiation attempts before aborting an event.

To optimize memory and execution output, verbose terminal logging was disabled (Next:numberShowInfo=0, Next:numberShowProcess=0, Next:numberShowEvent=0), while non-default settings were displayed upon initialization (Init:showChangedSettings=on).

In the analysis loop, prompt photons originating directly from the hard interaction were selected. Secondary photons from hadronic decays (e.g., $\pi^0 \to \gamma\gamma$) were rejected by applying status code filters:

1. Particle Identification: Photons were identified by PDG code 22 (pythia.event[i].id()==22).
2. Status Selection: Only outgoing particles from the hard process with absolute status code 23 (statusAbs()==23) were selected.
3. Kinematic Cuts: The selected prompt photons were required to fall within the primary fiducial volume defined by:

$$p_T \geq 0.5 \text{ GeV}/c, \quad -2.0 \leq y \leq 2.0,$$

where $p_T$ is the photon transverse momentum and $y$ is the rapidity.

For each accepted event, the differential distributions were accumulated into histograms weighted by the NLO event weight $w_i = \text{\texttt{pythia.info.weight()}}$ to ensure proper normalization to the total cross section $\sigma_{\text{gen}}$ in millibarns (mb). The constructed physical observables include:

5. Transverse momentum distribution ($p_T$) in the range $[0.5, 5.0]$ GeV$/c$,
6. Rapidity distribution ($y$) in the range $[-2.5, 2.5]$,
7. Polar scattering angle cosine ($\cos\theta = p_z/|\mathbf{p}|$) in the range $[-1.0, 1.0]$.

To compare differential cross sections at the parton level, calculated in the LO and NLO approximations, we used the relation $\hat{R} = \frac{d\hat{\sigma}_{NLO}(q\bar{q}\to g\gamma)}{d\hat{\sigma}_{LO}(q\bar{q}\to g\gamma)}$ and find the following expression $\hat{R} \approx 0.005\alpha_s$.

The hadron level relationship $R = \frac{d\sigma_{NLO}(q\bar{q}\to g\gamma)}{d\sigma_{LO}(q\bar{q}\to g\gamma)}$ is determined using PDF. Investigation of the dependence of the following ratios:

1. $R = \frac{d\sigma_{NLO}(q\bar{q}\to g\gamma)}{d\sigma_{LO}(q\bar{q}\to g\gamma)}$,
2. $R = \frac{d\sigma_{NLO_{POL}}(q\bar{q}\to g\gamma)}{d\sigma_{NLO}(q\bar{q}\to g\gamma)}$, and
3. $R = \frac{d\sigma_{\text{FeynCalc}}(q\bar{q}\to q\bar{q}\gamma)}{d\sigma_{\text{PYTHIA}}(q\bar{q}\to q\bar{q}\gamma)}$

on the sum energy of colliding particles $\sqrt{s}$, the transverse momentum $p_T$ of produced photons, the cosine of scattering angle $Cos(\theta)$, rapidity $y$ of photon and $x_T$ has been carried out.

## III. NUMERICAL RESULTS AND THEIR DISCUSSION

### 3.1 Differential cross-section of subprocess of annihilation of quark-antiquark pair $q\bar{q} \to g\gamma$ at collision of nonpolarized protons

Figure 1(a,b,c,d,e) shows the dependence of the differential cross section of the subprocess $q\bar{q} \to g\gamma$, calculated in the NLO, on the sum of the energy of colliding particles $\sqrt{s}$, the transverse momentum of prompt photons $p_T$, the cosine of the scattering angle $Cos(\theta)$ of prompt photons, the rapidity $y$ and $x_T$ at $\sqrt{s}$=10 GeV. In this case, restrictions are imposed on the transverse momentum $p_T \geq 0.5$ GeV/c and the rapidity of prompt photons $-2 \leq y \leq 2$.

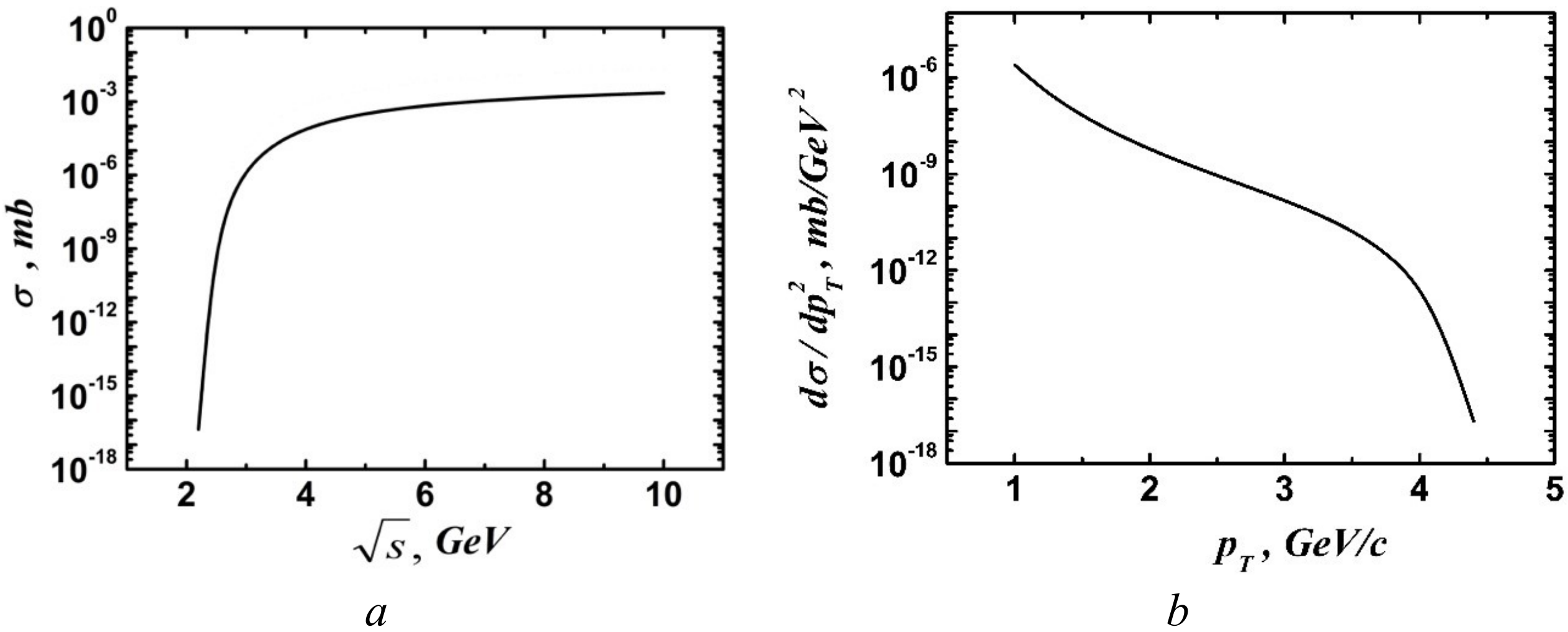

*a* *b*

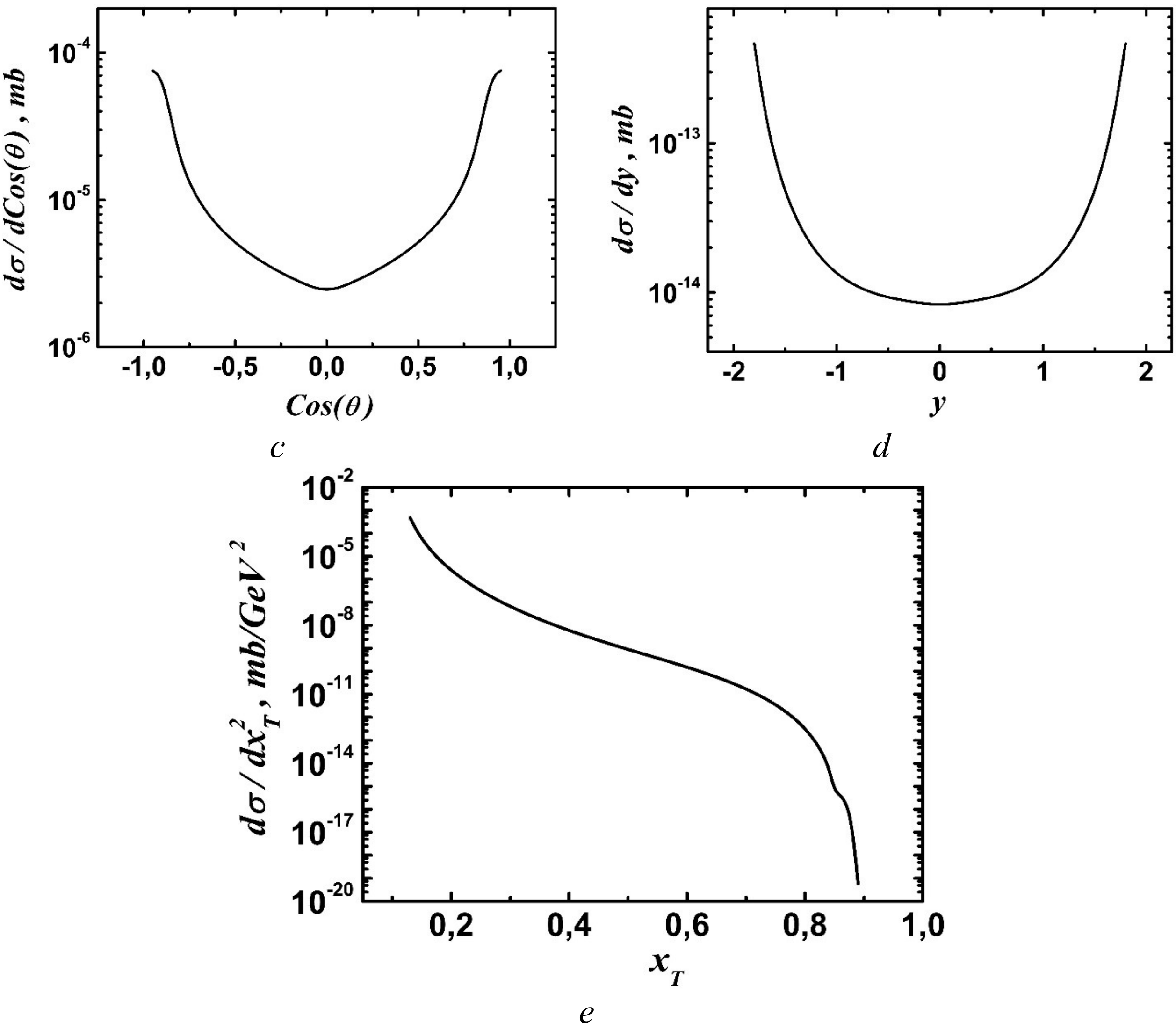

Figure 1(a,b,c,d,e). The dependence of the differential cross section of the subprocess $q\bar{q} \rightarrow g\gamma$, calculated in the NLO, on the sum of the energy of colliding particles $\sqrt{s}$ (a), the transverse momentum of prompt photons $p_T$ (b), the cosine of the scattering angle $Cos(\theta)$ of prompt photons (c), the rapidity $y$ (d) and $x_T$ (e) at $\sqrt{s}$=10 GeV.

It can be seen from Figure 1(a) that at low energies the total cross section increases rapidly with the increase in the value of $\sqrt{s}$. The total cross-section reaches a maximum at the value $\sqrt{s}$=5.2 GeV, and then slowly decreases with further growth of the value $\sqrt{s}$. We assume that at low energies, quarks and antiquarks have a small relative velocity and, therefore, strongly overlap in space and time. This increases the probability of gluon and photon emission, which have no mass and can carry most of the energy. The cross-section is proportional to the square of the electric charge of the quark. Consequently, the differential cross section for upper quarks is greater than for lower quarks. The differential cross-section is also suppressed by the running constant of the connection $\alpha_{ss}$ QCD, which decreases with the growth of energy. Therefore, the differential cross-section for higher energies is smaller than for lower energies.

As can be seen from Figure 1(b), the differential cross section $d\sigma/dp_T^2$ decreases sharply with increasing $p_T$. This is because the production of high-$p_T$. particles requires more energy and is suppressed by the running QCD coupling constant $\alpha_s$, which decreases with increasing energy scale. In addition, such a rapid decrease is characteristic of hard scattering processes such as quark-antiquark pair annihilation, due to the increase in phase space available for particles in the higher-momentum final state.

The dependence of the differential cross section $d\sigma/dCos(\theta)$ on the scattering angle cosine $Cos(\theta)$ has a symmetrical peaked shape with a maximum at $Cos(\theta) = \pm 1$ and a minimum at $Cos(\theta) = 0$ (Figure 1(c)). This is due to the fact that the $q\bar{q} \rightarrow g\gamma$ subprocess is dominated by the $t$-channel exchange of a virtual photon, which has spin 1 and an even-even nature. The angular distribution of the final state particles reflects the conservation of angular momentum and parity in the subprocess.

As shown in Figure 1(d), the dependence of the differential cross section on the photon rapidity y has a maximum at $y = \pm 1.95$ and decreases with decreasing absolute value of rapidity $y$. The differential cross section is minimal when $y = 0$.

Figure 1(e) shows that the $d\sigma/dx_T^2$ differential cross section decreases rapidly with increasing $x_T$, indicating suppression of gluon and photon production with large transverse momenta. This is consistent with theoretical predictions based on pQCD, which show that the cross section is proportional to $\alpha_s\alpha$ where $\alpha_s$ is the tight coupling constant and $\alpha$ is the fine structure constant. Since $\alpha_s$ and $\alpha$ are small, the cross section is expected to be small for large $x_T$. Figure 1(e) also shows that the dependence of the differential cross section on $x_T$ is not linear, but has a power-law character. This is due to the fact that the cross-section is sensitive to the quark and antiquark PDFs, which describe the probability of finding a parton with a given fraction of the proton momentum. The PDFs are not constant, but vary depending on the fraction of the momentum and the energy scale of the process. In general, the PDFs decrease with increasing momentum fraction, reflecting the fact that there are fewer partons with high momenta inside the proton.

### 3.2 Differential cross-section of subprocess of annihilation of quark-antiquark pair $q\bar{q} \rightarrow g\gamma$ at collision of longitudinally polarized protons

Figure 2(a,b,c,d,e) shows the dependences of the differential cross sections prompt photon production in annihilation of quark-antiquark pair $q\bar{q} \rightarrow g\gamma$ calculated taking into account the longitudinal polarization of colliding protons on the sum of the energy of colliding particles $\sqrt{s}$, the transverse momentum of prompt photons $p_T$, the cosine of the scattering angle of prompt photons, the rapidity y and $x_T$ at $\sqrt{s}$=10 GeV.

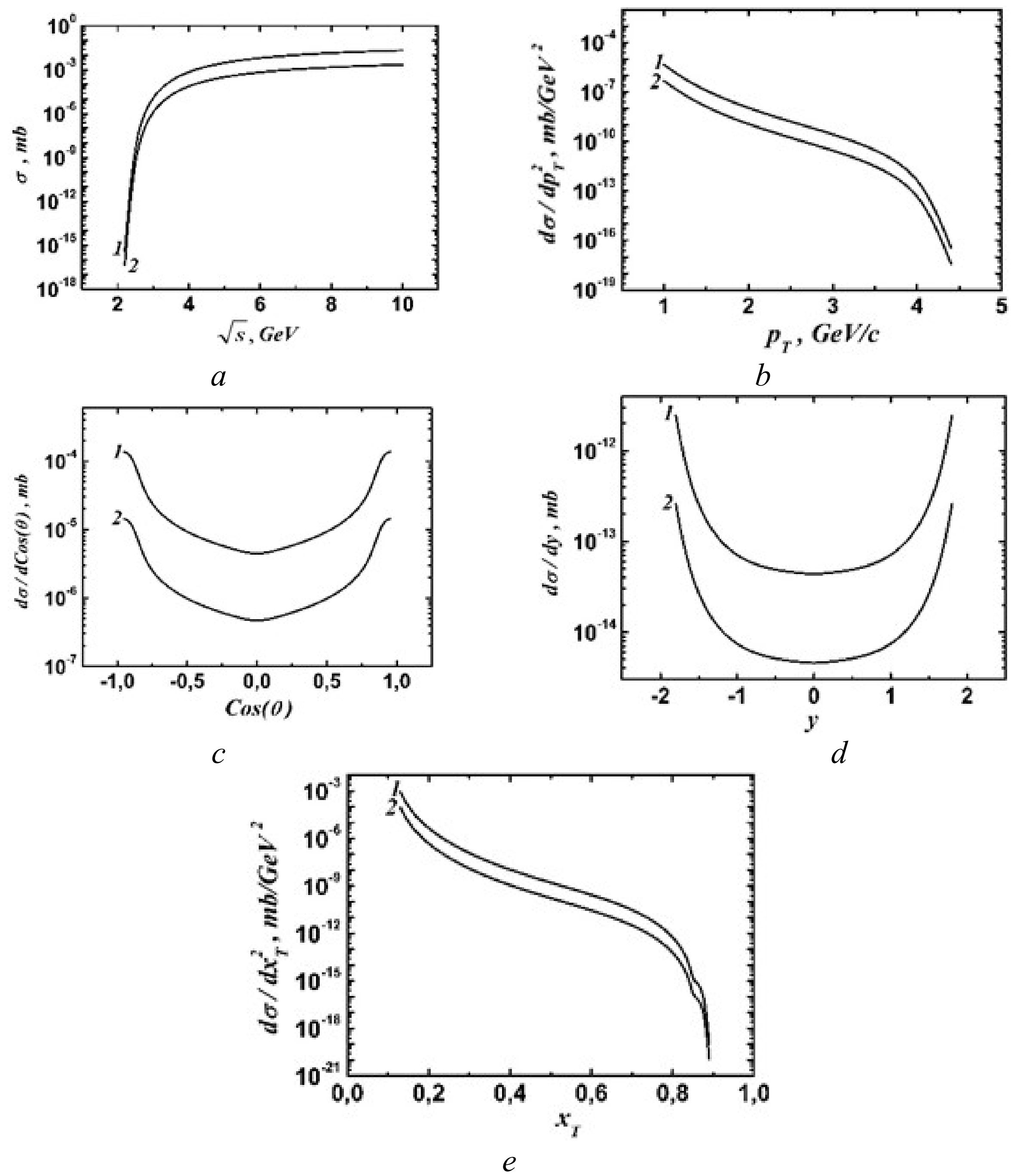

Figure 2(a,b,c,d,e). Dependence of the differential cross section of the subprocess $q\bar{q} \rightarrow g\gamma$ taking into account the polarization of colliding protons, calculated in the NLO at $P_1$, $P_2$=0.9, -0.9, that is $P_1P_2$=-0.81 (curve 1) and at $P_1$=$P_2$=±0.9, that is $P_1P_2$=0.81 (curve 2) on the sum of the energy of colliding particles $\sqrt{s}$ (a), the transverse momentum of prompt photons $p_T$ (b), the cosine of the scattering angle of prompt photons (c), the rapidity y (d) and $x_T$ (e) at $\sqrt{s}$=10 GeV

From the dependences in Figure 2(a,b,c,d,e) it is evident that polarization does not affect the character of the dependences. As can seen from Figure 2(a,b,c,d,e) with opposite polarization the differential cross-section increases, and with the same polarization it decreases, which is similar to the LO.

From figure 2(a) it is evident that in the considered range of changes in the energy of colliding protons, the total cross sections σ of the subprocess increase with an increase in the sum of the energies of the colliding particles $\sqrt{s}$.

The differential cross section of the subprocess decreases rapidly as the transverse momentum $p_T$ of prompt photons increases (Figure 2(b)).

The dependence of the differential cross section $d\sigma/dCos(\theta)$ on the cosine of the scattering angle $Cos(\theta)$ has a symmetrical pointed shape with a maximum at $Cos(\theta) = \pm 1$ and a minimum at $Cos(\theta) = 0$ (Figure 2(c)).

The differential cross-section has a maximum at $y = \pm 1.95$ and decreases with a change in $y$, reaching its minimum at the point $y = 0$ (Figure 2(d)).

Figure 3 shows a 3D graph of the dependence of the doublespin asymmetry $A_{LL}$ of the quark-antiquark pair annihilation subprocess calculated in the NLO on the polarization order $P_1$ and $P_2$ of the colliding particles.

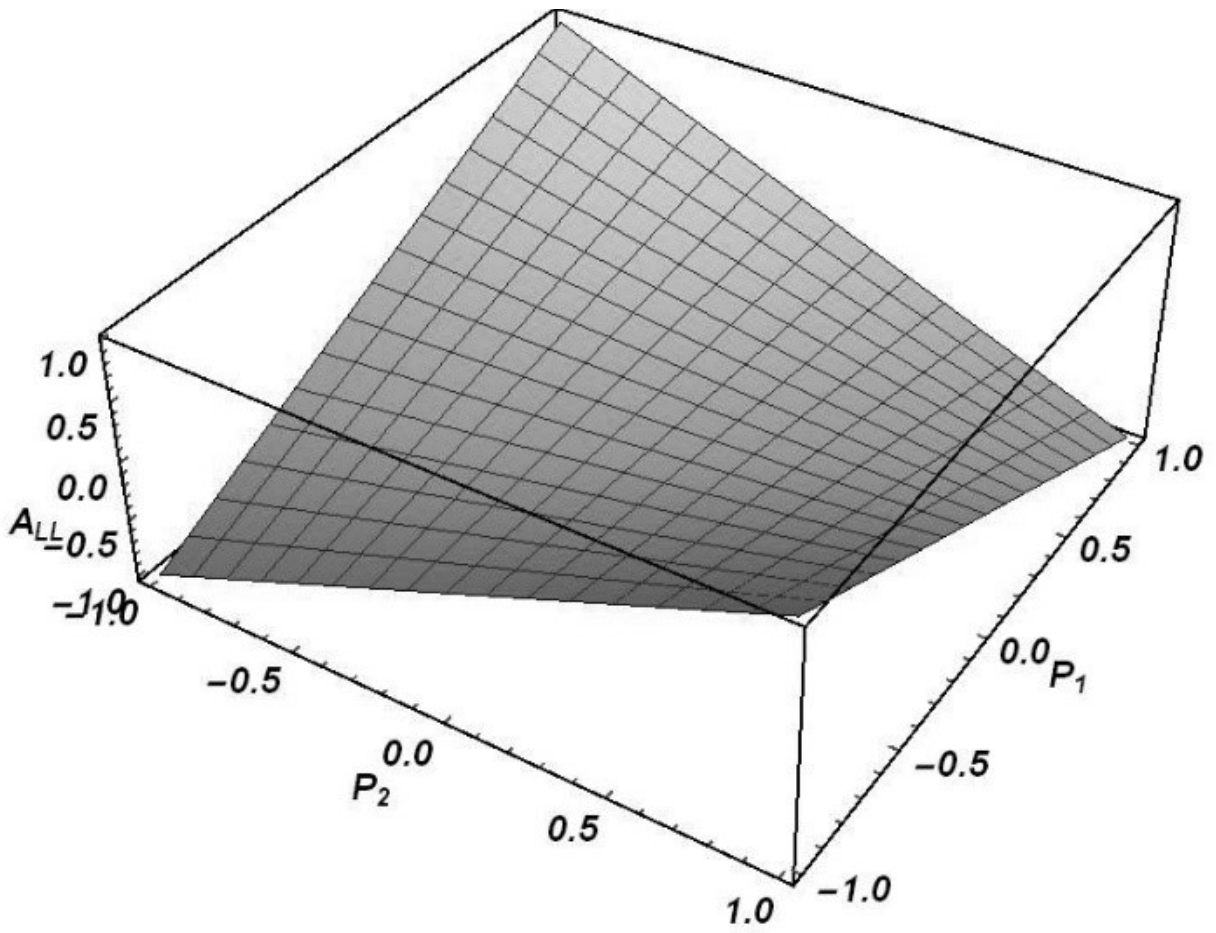

Figure 3. 3D graph of the dependence of the doublespin asymmetry $A_{LL}$ of the annihilation subprocess of the quark-antiquark pair $q\bar{q} \to g\gamma$ on the polarization order $P_1$ and $P_2$ of the colliding particles at $\sqrt{s}$=10 GeV.

Maximum positive asymmetry, $A_{LL} \approx +1.0$ is achieved when the polarizations $P_1$ and $P_2$ have the same sign and maximal magnitude (i.e., $P_1 = P_2 = 1$ or $P_1 = P_2 = -1$). Physically, this implies that the reaction cross-section $\sigma(q \uparrow \bar{q} \uparrow) \gg \sigma(q \uparrow \bar{q} \downarrow)$. In other words, the annihilation process is most probable when the quark and antiquark spins are aligned parallel to each other (either both along or both against their direction of motion). This is a direct consequence of the spin structure of the QCD and QED interaction vertices, specifically the helicity selection rules.

Maximum negative asymmetry, $A_{LL} \approx -1.0$ is observed when the polarizations have opposite signs ($P_1 = 1$, $P_2 = -1$, or $P_1 = -1$, $P_2 = 1$). In this scenario, the cross-section $\sigma(q \uparrow \bar{q} \downarrow) \gg \sigma(q \uparrow \bar{q} \uparrow)$, and the process favors an antiparallel spin configuration.

The asymmetry vanishes, $A_{LL}$=0 if at least one of the initial particles is nonpolarized ($P_1 = 0$ or $P_2 = 0$). This is a natural consequence, as the absence of polarization implies no preferred spin direction and, therefore, no possible asymmetry.

3.3 The dependences of differential cross sections of the subprocesses annihilation of quark-antiquark pair $q\bar{q} \rightarrow g\gamma$ on the sum of energies of colliding protons $\sqrt{s}$, transverse momentum $p_T$, cosine of scattering angle $Cos(\theta)$, rapidity y of photon and $x_T$ simulated in PYTHIA 8.316 + POWHEG BOX.

Figure 4(a,b,c,d,e) shows the dependences of the total and differential unpolarized cross sections for subprocess: $q\bar{q} \rightarrow g\gamma$ on the sum of energies of colliding protons $\sqrt{s}$, transverse momentum $p_T$, cosine of scattering angle $Cos(\theta)$, rapidity y of photon and $x_T$ in the PYTHIA 8.316 + POWHEG BOX simulation at $\sqrt{s}$ =10 GeV. These curves agree with the curves obtained from calculation in the FeynCalc, which run higher.

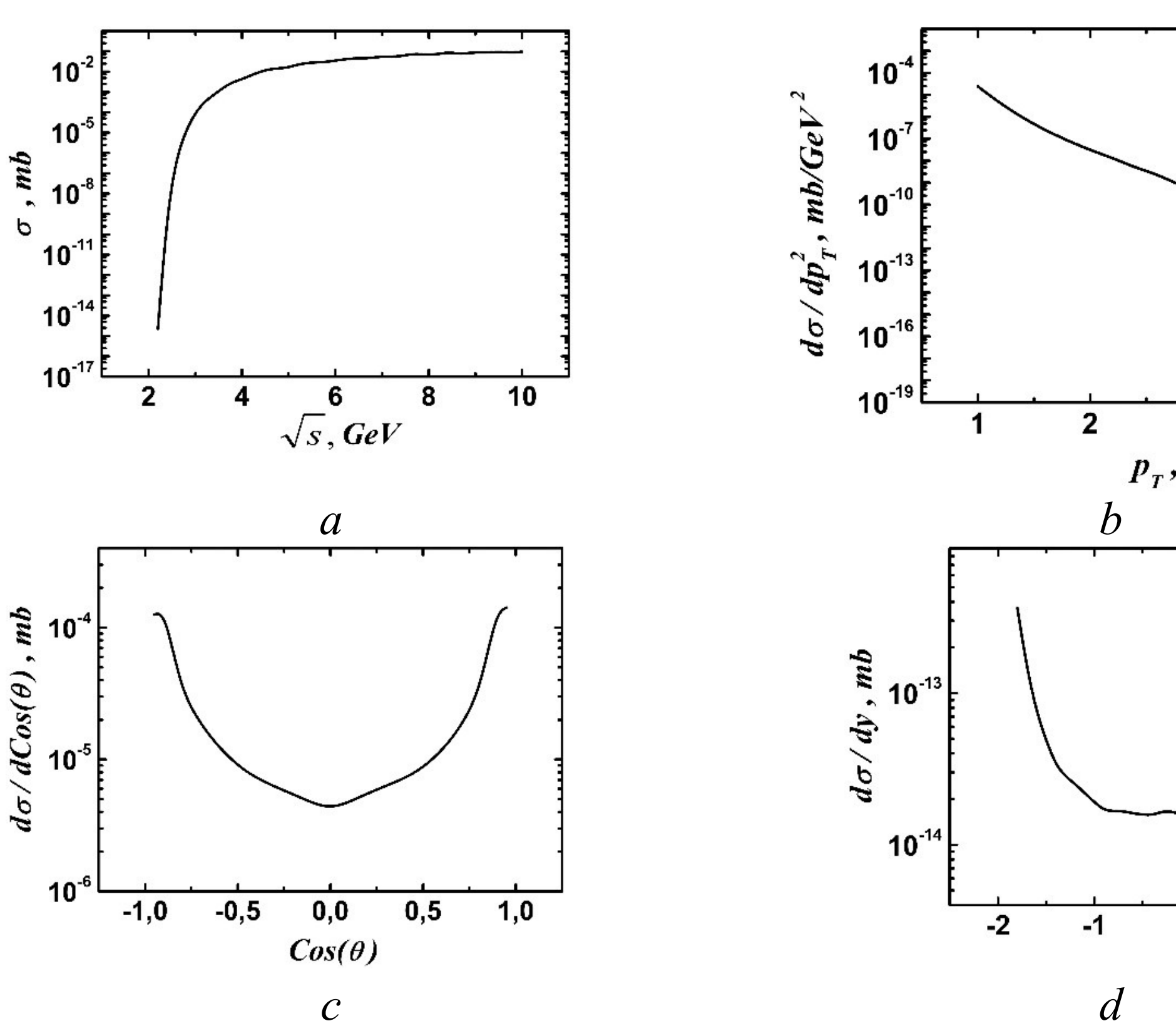

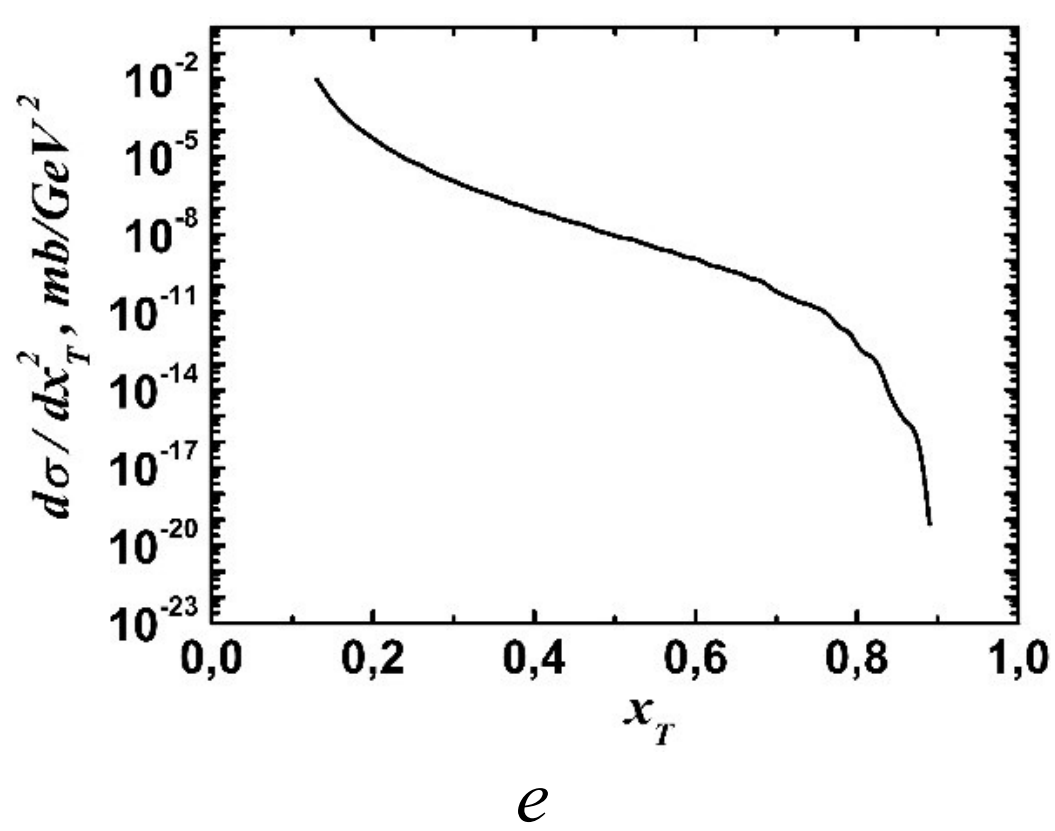

*e*

Figure 4(a,b,c,d,e). The dependence of the differential cross section of the subprocess $q\bar{q} \to g\gamma$, calculated in the NLO (PYTHIA 8.316 simulation), on the sum of the energy of colliding particles $\sqrt{s}$ (a), the transverse momentum of prompt photons $p_T$ (b), the cosine of the scattering angle $Cos(\theta)$ of prompt photons (c), the rapidity $y$ (d) and $x_T$ (e) at $\sqrt{s}$=10 GeV.

Figure 4(a,b,c,d,e) illustrates the differential cross-sections for the production of prompt photons via the subprocess $q\bar{q} \to g\gamma$ in proton-proton collisions at $\sqrt{s} = 10$ GeV (NICA kinematic regime). Crucially, the presented curves are not purely analytical fixed-order theoretical functions, but are the direct output of a sophisticated MC event generator framework. Specifically, the hard matrix elements were calculated at NLO using POWHEG-BOX, and then matched with PYTHIA 8.316 to simulate Parton Showers (PS), hadronization, and the full event kinematics.

This NLO+PS approach means that PYTHIA's algorithms directly shape the observed distributions by supplementing the hard scattering with multiple soft and collinear parton emissions, governed by Sudakov form factors, yielding a highly realistic physical picture.

The total cross-section exhibits a sharp threshold behavior near $\sqrt{s} \approx 2$ GeV, corresponding to the energy needed to produce a photon and a recoiling jet above the minimum fiducial cutoff ($p_T \geq 0.5$ GeV/$c$). As energy increases, PYTHIA samples partons at lower momentum fractions ($x$) from the CT14nlo PDFs. The steep exponential growth of gluon and sea-quark densities at low $x$ produces the plateau-like rise of the cross-section up to 10 GeV (Fig.4(a)).

The distributions $d\sigma/dp_T^2$ and $d\sigma/dx_T^2$ (where $x_T = 2p_T/\sqrt{s}$) show a steeply falling, power-law-like tail characteristic of perturbative QCD hard scattering (Fig.4(b,e)). However, the precise shape-especially at the lower $p_T$ boundary-is strongly modified by PYTHIA's Initial State Radiation (ISR). The parton shower gives the $q\bar{q}$ system an intrinsic transverse momentum "kick" prior to the hard collision, which smears the strictly back-to-back kinematics of the leading-order process and fills in the lower end of the spectrum dynamically.

The distinctive $U$-shape of $d\sigma/d\text{Cos}\theta$ peaks as $\text{Cos}\theta \to \pm 1$ (forward/backward) and has a minimum at $\text{Cos}\theta = 0$ (transverse) (Fig.4(c)). The hard NLO matrix element for $q\bar{q} \rightarrow g\gamma$ contains $t$- and $u$-channel quark propagators, creating collinear enhancements $\propto 1/(1 \mp \mathrm{Cos}\theta)$. In a purely analytical calculation, this approaches a divergence. In this PYTHIA simulation, the divergences are rigorously regulated by both the explicit $p_T$ generation cutoff (0.5 GeV/$c$) and PYTHIA's internal kinematic showering limits, leading to smooth, well-behaved physical peaks.

The rapidity distribution $d\sigma/dy$ mirrors the angular dependence (Fig.4(d)). The plateau at $y \approx 0$ reflects central, high-angle scattering. The peaks towards $y \rightarrow \pm 2$ reflect the fundamental preference of gauge theory processes for forward/backward scattering. The abrupt cutoff exactly at $|y| = 2.0$ demonstrates the strict fiducial phase-space filter applied dynamically during the PYTHIA 8.316 event analysis loop.

### 3.4 Comparison of differential cross sections annihilation of quark-antiquark pair $\boldsymbol{q\bar{q} \rightarrow g\gamma}$ calculated in LO and NLO at collision of nonpolarized and longitudinally polarized protons, simulated in PYTHIA 8.316

#### 3.4.1 Investigation of ratio $R = \frac{d\sigma_{NLO}(q\bar{q}\rightarrow g\gamma)}{d\sigma_{LO}(q\bar{q}\rightarrow g\gamma)}$

The contribution of the NLO to the differential cross section was determined by studying the dependence of the ratio $R = \frac{d\sigma_{NLO}(q\bar{q}\rightarrow g\gamma)}{d\sigma_{LO}(q\bar{q}\rightarrow g\gamma)}$ on the sum of the energy of colliding particles $\sqrt{s}$ and the transverse momentum of prompt photons $p_T$ at $\sqrt{s}$=10 GeV.

Figure 5(a,b) shows the dependence of the ratio $R = \frac{d\sigma_{NLO}(q\bar{q}\rightarrow g\gamma)}{d\sigma_{LO}(q\bar{q}\rightarrow g\gamma)}$ on the sum of energy of colliding protons $\sqrt{s}$ and the transverse momentum of prompt photons $p_T$ at $\sqrt{s}$=10 GeV.

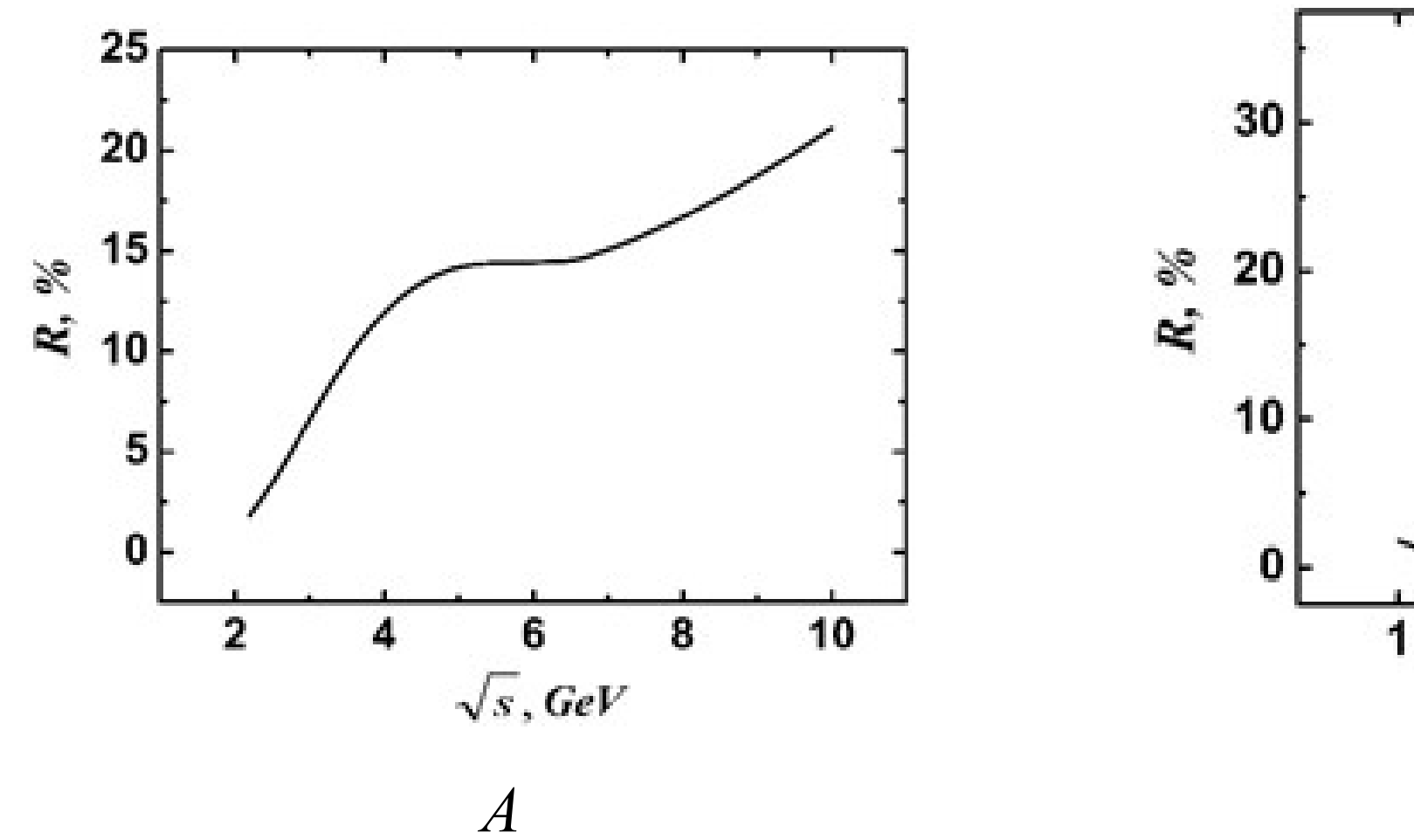

*A*

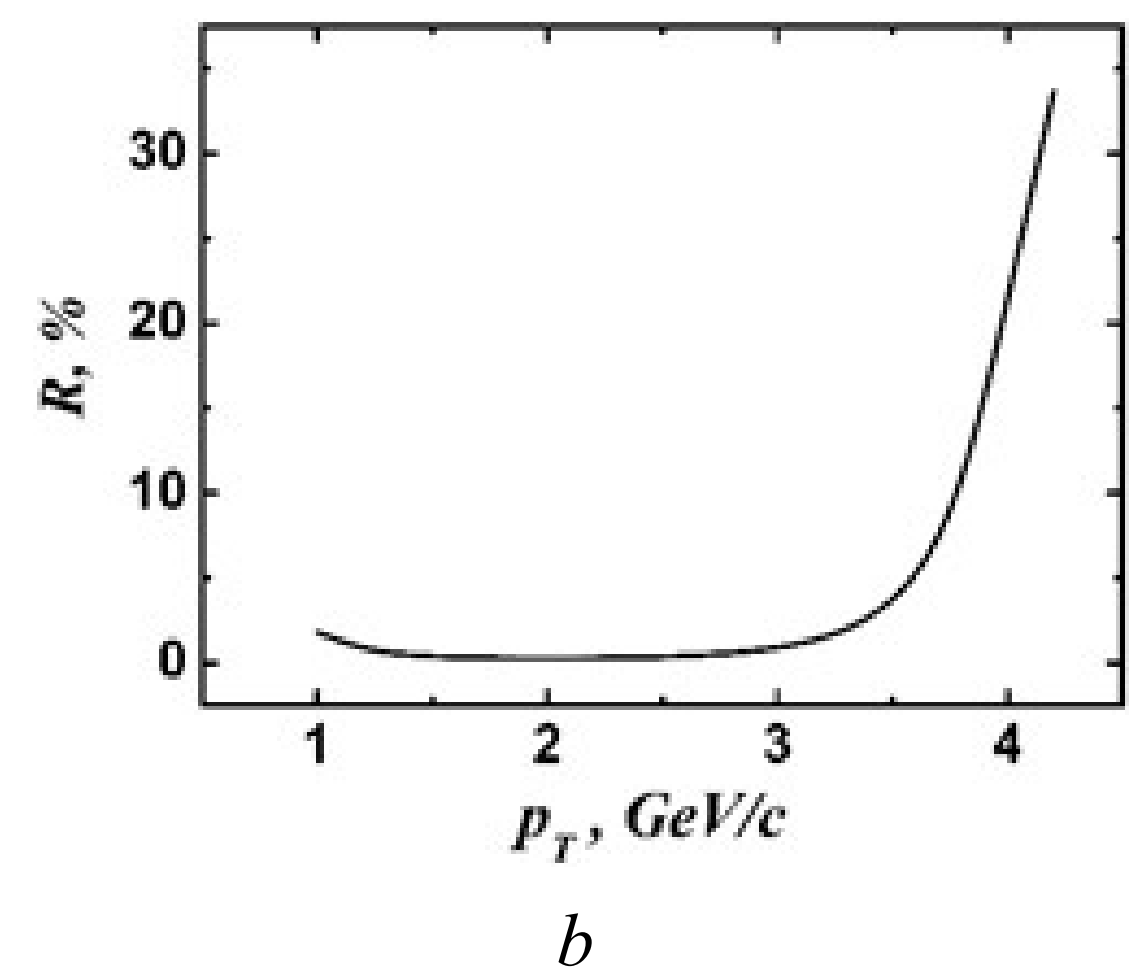

*b*

Figure 5(a,b). Dependence of the ratio $R = \frac{d\sigma_{NLO}(q\bar{q}\rightarrow g\gamma)}{d\sigma_{LO}(q\bar{q}\rightarrow g\gamma)}$ on the sum of energy of colliding particles $\sqrt{s}$ (a), transverse momentum of promptt photons $p_T$ (b) at

$\sqrt{s}$=10 GeV

As can be seen from Figure 5(a), as the value of $\sqrt{s}$ increases in the [2. 10] GeV range, the value of R increases, which shows the importance of taking into account the NLO calculations at high energies of $\sqrt{s}$.

As can seen from Figure 5(b), the dependence of $R$ on the transverse momentum of prompt photons $p_T$ indicate that the contribution of the NLO corrections reaches up to 30% at large transverse momenta of prompt photons $p_T$.

Comparing the contributions of the LO and NLO calculations, one can see that the LO contribution is proportional to $\alpha_s\alpha$, while the NLO contribution is proportional to $\alpha_s^2\alpha$. NLO calculations include additional Feynman diagrams and higher-order corrections. Increasing the number of Feynman diagrams in the NLO means including more complex and virtual particle exchanges, which leads to a more accurate description of the physical process. The interference term is proportional to $\alpha_s^2\alpha\, log(s/M^2)$, which increase as $p_T$ increases, where $M$ is the renormalization scale. The interference term changes sign at $s \approx M^2$, leading to a minimum in the cross section. The cross-section also increases as the values of the PDF increase at small values of $x$. Therefore, NLO calculations can prodece a differential cross-section contribution that is smaller than that predicted by the LO.

### 3.4.2 Investigation of ratios $R = \frac{d\sigma_{LO_{POL}}(q\bar{q}\to g\gamma)}{d\sigma_{LO}(q\bar{q}\to g\gamma)}$ and $\frac{d\sigma_{NLO_{POL}}(q\bar{q}\to g\gamma)}{d\sigma_{NLO}(q\bar{q}\to g\gamma)}$

Figure 6(a,b) show the ratios $R = \frac{d\sigma_{LO_POL}(q\bar{q}\to g\gamma)}{d\sigma_{LO}(q\bar{q}\to g\gamma)}$ and $\frac{d\sigma_{NLO_POL}(q\bar{q}\to g\gamma)}{d\sigma_{NLO}(q\bar{q}\to g\gamma)}$ on the sum of the energies of the colliding particles $\sqrt{s}$.

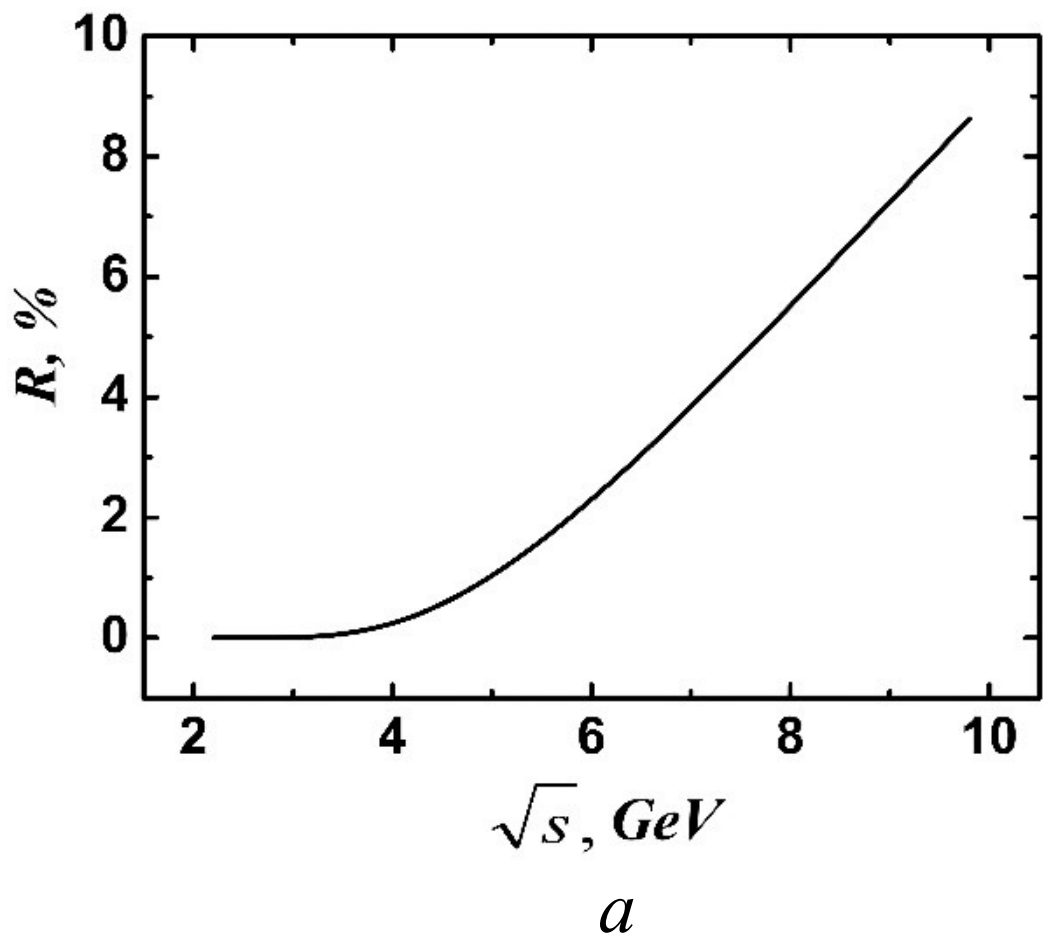

*a*

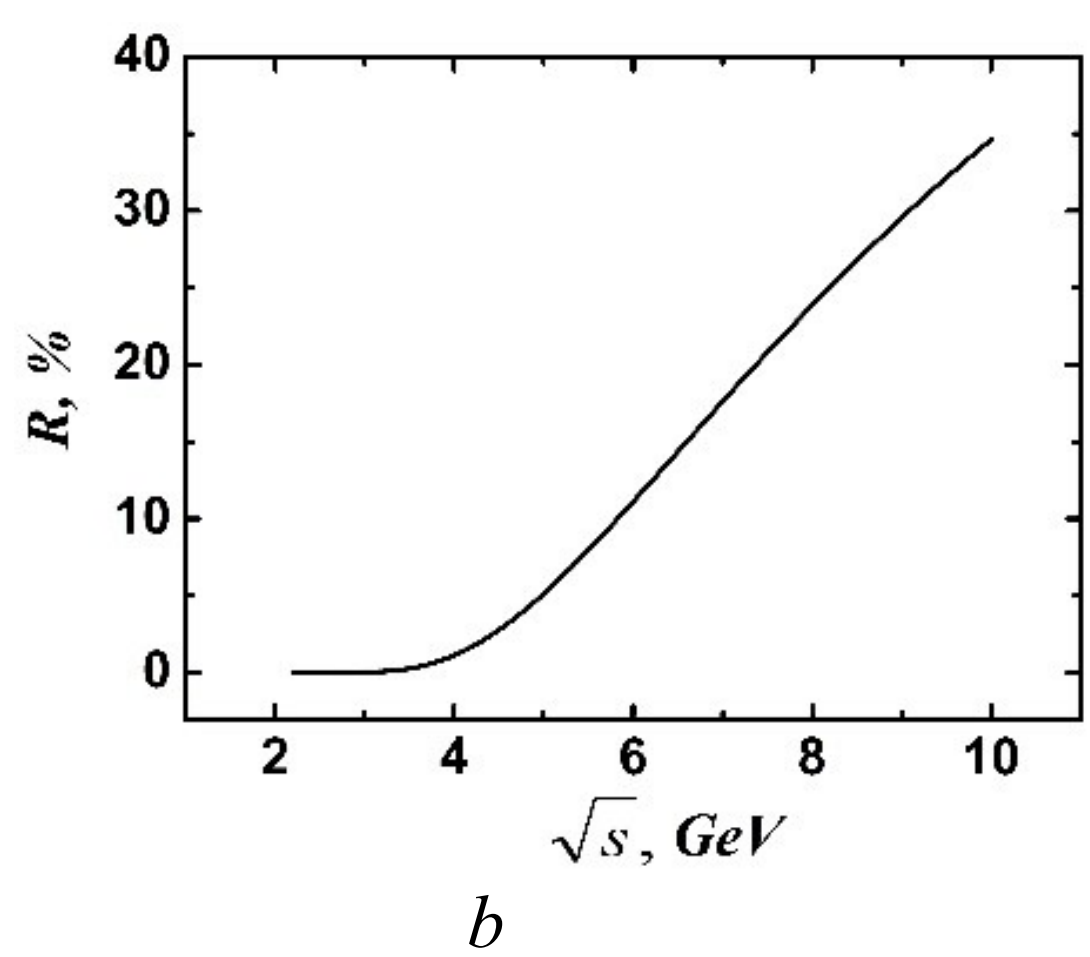

*b*

Figure 6(a,b). Dependence of the ratios $R = \frac{d\sigma_{LO_POL}(q\bar{q}\to g\gamma)}{d\sigma_{LO}(q\bar{q}\to g\gamma)}$ (a) and $R = \frac{d\sigma_{NLO_POL}(q\bar{q}\to g\gamma)}{d\sigma_{NLO}(q\bar{q}\to g\gamma)}$ (b) on the sum of the energy of colliding particles $\sqrt{s}$

As can be seen from Figure 6(a,b) as the sum energy of colliding particles increases $\sqrt{s}$, an increase in the effect of polarization in the LO and NLO is observed. It can be seen from the differential cross sections of the LO and NLO for the quark-antiquark pair annihilation subprocess $q\bar{q} \rightarrow g\gamma$ that the longitudinally polarization of proton strongly influences the NLO than the LO. The contribution of the calculation to the NLO on the differential cross section is significant at high energies of the colliding particles.

The reason for this may be: firstly, new Feynman diagrams appear in NLO calculations that include corrections for the virtual loop. These new diagrams often produce contributions with opposite signs compared to the LO diagrams, resulting in a decrease in the overall cross-section of the process. However, for specific observable quantities such as polarization asymmetry, this suppression may not be as effective, leading to a stronger relative influence of the remaining polarization-sensitive terms. Second, NLO calculations account for additional kinematic configurations compared to LO, including higher order parton scattering and gluon emission. These configurations may be more sensitive to proton spin orientation, resulting in greater polarization dependence compared to the more constrained kinematic pattern in LO.

### 3.4.3 Investigation of ratio $R = \frac{d\sigma_{\text{FeynCalc}}(q\bar{q}\rightarrow g\gamma)}{d\sigma_{\text{PYTHIA}}(q\bar{q}\rightarrow g\gamma)}$

Investigation of ratio $R = \frac{d\sigma_{\text{FeynCalc}}(q\bar{q}\rightarrow g\gamma)}{d\sigma_{\text{PYTHIA}}(q\bar{q}\rightarrow g\gamma)}$ allowed to quantify the impact of higher-order logarithmic resummation, initial-state radiation (ISR), scale choices, and non-perturbative hadronization relative to pure fixed-order perturbation theory. $R = \frac{d\sigma_{\text{FeynCalc}}(q\bar{q}\rightarrow g\gamma)}{d\sigma_{\text{PYTHIA}}(q\bar{q}\rightarrow g\gamma)}$ evaluated across identical kinematic observables at $\sqrt{s} = 10$ GeV.

The dependence of the ratio $R$, defined as the differential cross-section calculated analytically using FeynCalc (fixed-order perturbation theory) divided by the results obtained from the PYTHIA 8.316 NLO+PS simulation.

While analytical calculations provide the exact matrix elements for the isolated $2 \rightarrow 2$ hard subprocess ($q\bar{q} \rightarrow g\gamma$), they fail to capture the complex, multi-particle dynamics of a real proton-proton collision. PYTHIA, conversely, supplements the hard scattering with Initial and Final State Radiation (ISR/FSR), Sudakov resummation, and strict energy-momentum conservation across the entire event. The ratio $R$ thus serves as a direct measure of the impact of parton showering and multi-gluon emissions on the kinematic observables.

The ratio $R$ starts very low ($\approx 16\%$ at $\sqrt{s} \approx 2$ GeV) and monotonically approaches 100% ($R \approx 1$ at 10 GeV). At low energies near the kinematic threshold, the phase space is highly restricted. In PYTHIA, energy carried away by ISR soft gluons often shifts the prompt photon's momentum below the fiducial cut ($p_T \geq 0.5$ GeV/$c$), depleting the observable cross-section compared to the idealized analytical formula. As $\sqrt{s}$ increases, the available phase space broadens significantly. The hard scattering becomes the dominant factor, and the inclusive NLO+PS result converges with the fixed-order analytical calculation.

The ratio shows a dramatic growth from $\approx 10\%$ at low $p_T$ to higher fluctuating values ($> 60 - 80\%$) at large $p_T$). This highlights a fundamental limitation of fixed-order calculations. At low $p_T$ (the soft/collinear limit), analytical cross-sections suffer from logarithmic divergences. PYTHIA resolves this via Sudakov resummation, effectively resumming multiple soft gluon emissions. The ISR gives the $q\bar{q}$ system an intrinsic transverse momentum "kick," dynamically populating the low-$p_T$ spectrum and making the PYTHIA cross-section much larger than the bare analytical one ($R \ll 1$). At high $p_T$, the hard matrix element takes over, and the ratio climbs closer to unity.

Interestingly, the ratio $R(\text{Cos}\theta)$ remains relatively flat and oscillating around $50\% - 60\%$ across the entire angular range. The angular distribution is governed by the $t$- and $u$-channel quark propagators of the underlying gauge theory, which create the characteristic $U$-shape. While the parton shower heavily smears the energy ($p_T$) of the particles, it does not fundamentally alter the spin and helicity structure of the $q\bar{q} \to g\gamma$ matrix element. Therefore, the angular shape is preserved in both approaches, resulting in a roughly constant ratio.

The ratio exhibits a pronounced $U$-shape: it is lowest at central rapidities ($y \approx 0,\ R \approx 54\%$) and rises above 100 At central rapidities, PYTHIA's ISR redistributes the momentum, shifting events to wider rapidities. However, at the extreme forward/backward edges, the fixed-order analytical calculation overestimates the cross-section because it assumes the incoming partons possess their full initial momentum. PYTHIA, on the other hand, enforces strict global energy conservation; the energy depleted by the parton shower limits how far forward/backward the final state particles can be produced, dampening the PYTHIA cross-section at the edges and causing $R$ to spike.

The PYTHIA simulation shows the main characteristics of pQCD quite accurately. The transverse momentum ($p_T, x_T$) spectra exhibit the expected steeply falling power-law tails. The angular ($\text{Cos}\theta$) and rapidity ($y$) distributions display a characteristic $U$-shape, reflecting the collinear enhancements driven by the $t$- and $u$-channel quark propagators inherent to the hard gauge interactions.

## CONCLUSIONS

Presented calculations showed the significant of NLO calculations for the more accurate description of prompt photon production in annihilation of quark-antiquark pair $q\bar{q} \to g\gamma$ in proton-proton collision.

The dependence of differential cross-section on sum of energy of colliding protons $\sqrt{s}$ calculated in NLO is significant at large values of energy $\sqrt{s}$.

The difference between the differential cross-sections calculated at LO and NLO is small at low and high transverse momentum $p_T$.

The photon has a high probability of produced along the collision axis of proton (angles are 16 and 164 degrees).

From the dependences of ratio $R = \frac{d\sigma_{NLO}}{d\sigma_{LO}}$ on the sum of the energies of the colliding initial particles $\sqrt{s}$, the transverse momentum $p_T$ of the prompt photons, the cosine of the scattering angle *Cos(θ)*, the rapidity $y$ and $x_T$of the prompt photons it can be seen that the NLO corrections increasingly affect the description of the process. The differential cross-sections of annihilation of quark-antiquark pairs $q\bar{q} \to g\gamma$ calculated by the NLO are about 15% of the calculations performed in the LO. This corresponds to pQCD, in which the higher-order terms give larger radiative corrections at higher energies of the partons.

The longitudinal polarization of colliding protons does not change the nature of the dependence of the differential cross sections of the subprocesses annihilation of quark-antiquark pair $q\bar{q} \to g\gamma$ on kinematic parameters. The longitudinal polarization of colliding particles affects the values of the differential cross sections of quark-antiquark pair annihilation. The same polarization direction of colliding particles, the differential cross section of the subprocesses of annihilation of quark-antiquark pairs $q\bar{q} \to g\gamma$ decreases. Polarization of colliding particles strongly influences the subprocess of annihilation of quark-antiquark pairs $q\bar{q} \to g\gamma$. The influence of proton polarization on the differential cross section is indicated by the interaction of parton spins.

The inclusion of longitudinal polarization of the colliding protons affects the NLO calculations much more strongly than the LO calculations. This enhanced sensitivity arises because NLO incorporates additional virtual loop corrections and higher-order parton scattering configurations that exhibit a stronger dependence on the initial proton spin orientations compared to the strictly constrained kinematics of LO diagrams.

Polarization is significant at small $p_T$ for annihilation of quark-antiquark pair $q\bar{q} \to g\gamma$.

Annihilation of a quark-antiquark pair $q\bar{q} \to g\gamma$ is a strong process associated with color charge. The influence of polarization during annihilation is due to the conservation of angular momentum.

The calculation of the double spin asymmetry $A_{LL}$ provides a clear reflection of QCD and QED helicity selection rules. The asymmetry reaches a maximum positive value ($A_{LL} \approx +1.0$) when the spins of the colliding quarks and antiquarks are aligned parallel, indicating the highest probability for annihilation. Conversely, a maximum negative asymmetry ($A_{LL} \approx -1.0$) occurs for antiparallel spin configurations. As expected, the asymmetry strictly vanishes if either of the initial protons is nonpolarized.

The comparison between analytical and PYTHIA 8.316 simulation results reveals the fundamental limitations of pure fixed-order calculations at low transverse momenta. While analytical cross-sections suffer from logarithmic divergences in the soft/collinear limit, the PYTHIA simulation successfully regulates these via Sudakov resummation. Initial State Radiation (ISR) provides a dynamic transverse momentum "kick" to the colliding partons, significantly populating the low-$p_T$ region and preventing unphysical divergences.

Comparison of our calculations in NICA energies with NLO calculations carried out at the LHC and the American Tevatron energies showed that the NLO is approximately 15% and 35% of LO for the NICA and LHC, American Tevatron energies, correspondingly.

The results obtained indicate the need to take into account NLO contributions when simulation and analyzing experimental data on the process of prompt photon production in proton-proton collisions at NICA and FAIR energies.

## APPENDIX.
## THE FEYNMAN DIAGRAMS AND MATRIX ELEMENTS OF NEXT-TO-LEADING ORDER SUBPROCESS $q\bar{q} \to g\gamma$

Figure A(a,b,c,d,e,f,g,h,i,j,k,e,m)) shows the Feynman diagrams of subprocess $q\bar{q} \to g\gamma$ in NLO approximation.

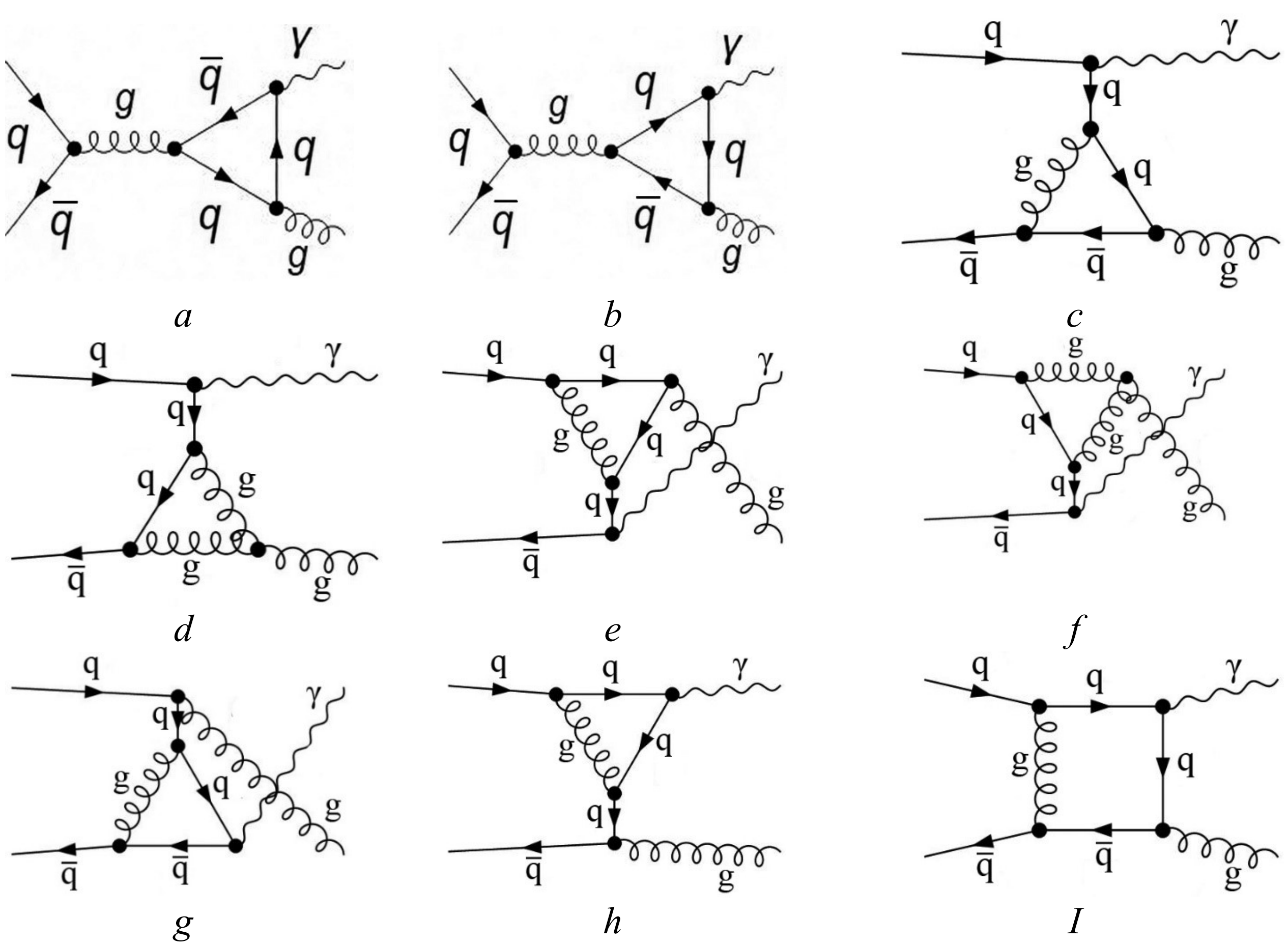

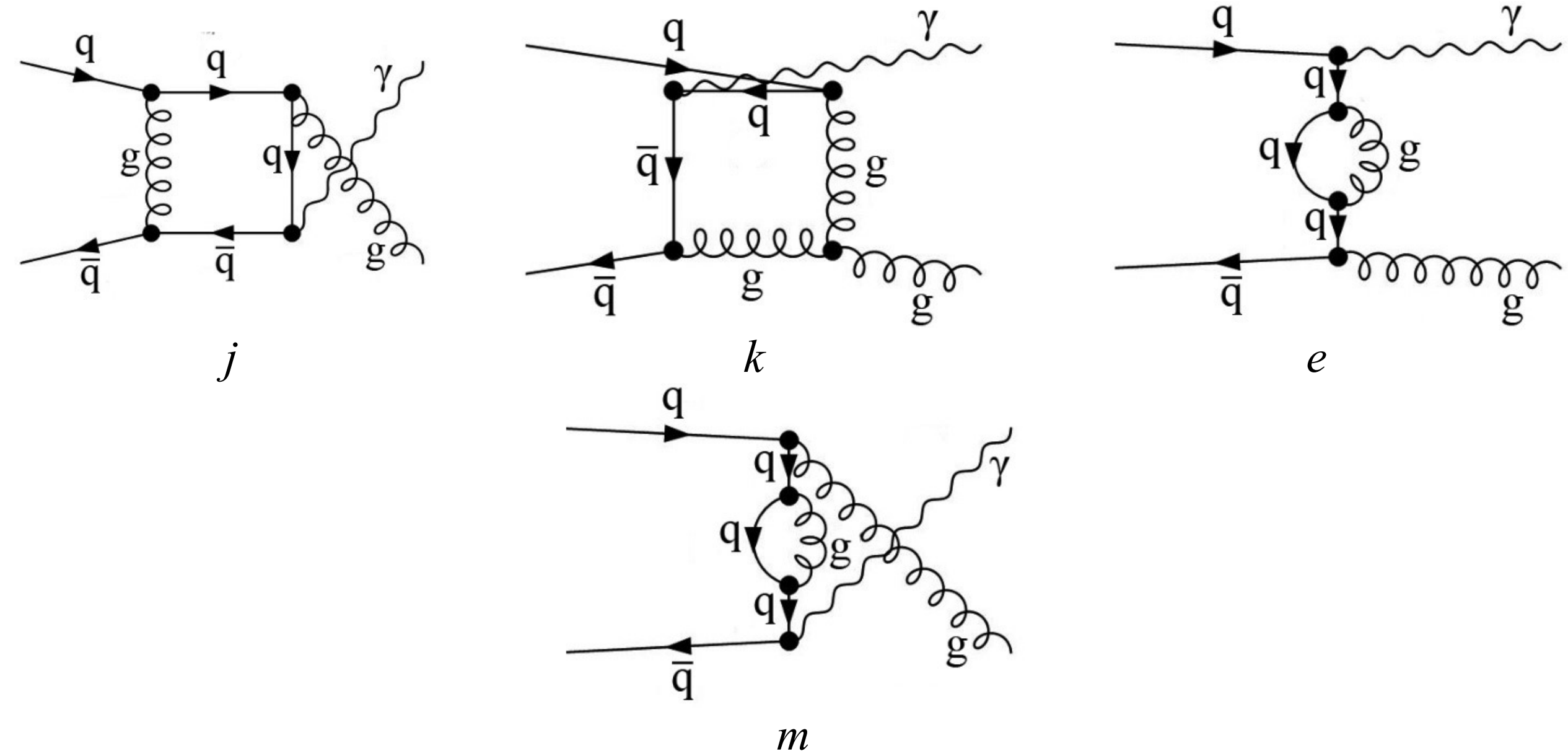

Figure A(a,b,c,d,e,f,g,h,i,j,k,l,m). Feynman diagrams of the one-loop corrections to the subprocess of annihilation of quark-antiquark pair $q\bar{q} \to g\gamma$.

In the parton level the Mandelstam invariants of subprocess $q(p_1) + \bar{q}(p_2) \to g(k_1) + \gamma(k_2)$ in LO and NLO is same:

$$\hat{s} = (p_1 + p_2)^2, \quad \hat{t} = (p_1 - k_1)^2 \quad \text{and} \quad \hat{u} = (p_2 - k_1)^2 .$$

Matrix elements of processes are:

$$M_1 = \frac{-iee_q g_s^3 T^{Glu5}_{Col2Col1} T^{Glu4}_{Col5Col6} T^{Glu5}_{Col6Col5} \bar{v}(p_2)\gamma^{\mu} u(p_1)(\hat{k}_2+\hat{q}+m_u)\gamma_{\mu}(\hat{q}-\hat{k}_1+m_u)\hat{\varepsilon}(k_1)(\hat{q}+m_u)\hat{\varepsilon}(k_2)}{144\pi^4(k_1+k_2)^2(q^2-m_u^2)((k_2+q)^2-m_u^2)((q-k_1)^2-m_u^2)},$$

$$M_2 = \frac{-iee_q g_s^3 T^{Glu5}_{Col2Col1} T^{Glu4}_{Col6Col5} T^{Glu5}_{Col5Col6} \bar{v}(p_2)\gamma^{\mu} u(p_1)(\hat{k}_2+\hat{q}+m_u)\gamma_{\mu}(\hat{q}-\hat{k}_1+m_u)\hat{\varepsilon}^*(k_1)(\hat{q}+m_u)\hat{\varepsilon}^*(k_2)}{144\pi^4(k_1+k_2)^2(q^2-m_u^2)((k_2+q)^2-m_u^2)((q-k_1)^2-m_u^2)},$$

$$M_3 = \frac{-iee_q g_s^3 T^{Glu5}_{Col6Col1} T^{Glu5}_{Col2Col5} T^{Glu4}_{Col5Col6} \bar{v}(p_2)\gamma\hat{q}\hat{\varepsilon}^*(k_1)(\hat{k}_1-\hat{q})\gamma(\hat{k}_1-\hat{p}_2)\hat{\varepsilon}^*(k_1)u(p_1)}{144\pi^4 q^2(p_2-k_1)^2(q-p_2)^2(q-k_1)^2},$$

$$M_4 = \frac{ee_q g_s^3 T^{Glu6}_{Col5Col1} T^{Glu5}_{Col2Col5} f^{Glu4Glu5Glu6}\left(k_1\varepsilon^*(k_1) - 2q\varepsilon^*(k_1)\right)\bar{v}(p_2)\gamma(\hat{q}-\hat{p}_2)\gamma\left(\hat{k}_1+\hat{p}_2\right)\hat{\varepsilon}^*(k_2)u(p_1)}{144\pi^4 q^2(p_2-k_1)^2(q-p_2)^2(q-k_1)^2} +$$

$$+\frac{ee_q g_s^3 T^{Glu6}_{Col5Col1} T^{Glu5}_{Col2Col5} f^{Glu4Glu5Glu6}\bar{v}(p_2)\left(\hat{q}-2\hat{k}_1\right)(\hat{q}-\hat{p}_2)\hat{\varepsilon}^*(k_1)\left(\hat{k}_1-\hat{p}_2\right)\hat{\varepsilon}^*(k_2)u(p_1)}{144\pi^4 q^2(p_2-k_1)^2(q-p_2)^2(q-k_1)^2} +$$

$$+\frac{ee_q g_s^3 T^{Glu6}_{Col5Col1} T^{Glu5}_{Col2Col5} f^{Glu1Glu5Glu6}\bar{v}(p_2)\hat{\varepsilon}^*(k_1)(\hat{q}-\hat{p}_2)\left(\hat{k}_1+\hat{q}\right)\left(\hat{k}_1-\hat{p}_2\right)\hat{\varepsilon}^*(k_2)u(p_1)}{144\pi^4 q^2(p_2-k_1)^2(q-p_2)^2(q-k_1)^2},$$

$$M_5 = \frac{iee_q g_s^3 T^{Glu5}_{Col5Col1} T^{Glu5}_{Col2Col6} T^{Glu4}_{Col6Col5} \bar{v}(p_2)\hat{\varepsilon}^*(k_2)\left(\hat{k}_2-\hat{p}_2\right)\gamma\left(\hat{q}-\hat{k}_1\right)\hat{\varepsilon}^*(k_1)\hat{q}\gamma u(p_1)}{144\pi^4 q^2(p_2-k_2)^2(q-p_2)^2(q-k_1)^2(p_2-k_1-k_2+q)^2},$$

$$M_6 = \frac{ee_q g_s^3 T^{Glu5}_{Col5Col1} T^{Glu6}_{Col2Col5} f^{Glu4Glu5Glu6}\left(k_1\varepsilon^*(k_1)-2q\varepsilon^*(k_1)\right)\bar{v}(p_2)\hat{\varepsilon}^*(k_2)\left(\hat{k}_2-\hat{p}_2\right)\gamma\left(\hat{k}_1+\hat{k}_2-\hat{p}_2-\hat{q}\right)\gamma u(p_1)}{144\pi^4 q^2(p_2-k_2)^2(q-k_1)^2(p_2+q-k_1-k_2)^2} +$$

$$+\frac{ee_q g_s^3 T^{Glu6}_{Col5Col1} T^{Glu5}_{Col2Col5} f^{Glu4Glu5Glu6}\bar{v}(p_2)\left(\hat{q}-2\hat{k}_1\right)(\hat{q}-\hat{p}_2)\hat{\varepsilon}^*(k_1)\left(\hat{k}_1-\hat{p}_2\right)\hat{\varepsilon}^*(k_2)u(p_1)}{144\pi^4 q^2(p_2-k_2)^2(q-k_1)^2(p_2+q-k_1-k_2)^2} +$$

$$+\frac{ee_q g_s^3 T^{Glu6}_{Col5Col1} T^{Glu5}_{Col2Col5} f^{Glu1Glu5Glu6}\bar{v}(p_2)\hat{\varepsilon}^*(k_1)(\hat{q}-\hat{p}_2)\left(\hat{k}_1+\hat{q}\right)\left(\hat{k}_1-\hat{p}_2\right)\hat{\varepsilon}^*(k_2)u(p_1)}{144\pi^4 q^2(p_2-k_2)^2(q-k_1)^2(p_2+q-k_1-k_2)^2},$$

$$M_7 = \frac{-iee_q g_s^3 T^{Glu4}_{Col5Col1} T^{Glu4}_{Col2Col6} T^{Glu5}_{Col6Col5} \bar{v}(p_2)\gamma\hat{q}\hat{\varepsilon}^*(k_2)\left(\hat{k}_2-\hat{q}\right)\gamma\left(\hat{k}_2-\hat{p}_2\right)\hat{\varepsilon}^*(k_1)u(p_1)}{144\pi^4 q^2(p_2-k_2)^2(q-p_2)^2(q-k_2)^2},$$

$$M_8 = \frac{iee_q g_s^3 T^{Glu5}_{Col6Col1} T^{Glu4}_{Col2Col5} T^{Glu5}_{Col5Col6} \bar{v}(p_2)\hat{\varepsilon}^*(k_1)\left(\hat{k}_1-\hat{p}_2\right)\gamma\left(\hat{q}-\hat{k}_2\right)\hat{\varepsilon}^*(k_2)\hat{q}\gamma u(p_1)}{144\pi^4 q^2(p_2-k_1)^2(q-k_2)^2(p_2+q-k_1-k_2)^2},$$

$$M_9 = \frac{-iee_q g_s^3 T_{Col5Col1}^{Glu5} T_{Col2Col6}^{Glu5} T_{Col6Col5}^{Glu4} \bar{\upsilon}(p_2)\gamma(\hat{q}+\hat{p}_2)\hat{\varepsilon}^*(k_1)(\hat{k}_1-\hat{p}_2-\hat{q})\hat{\varepsilon}^*(k_2)(\hat{k}_1+\hat{k}_2-\hat{p}_2-\hat{q})\gamma u(p_1)}{144\pi^4 q^2(p_2+q)^2(q+p_2-k_1)^2(p_2+q-k_1-k_2)^2},$$

$$M_{10} = \frac{-iee_q g_s^3 T_{Col5Col1}^{Glu5} T_{Col2Col6}^{Glu5} T_{Col6Col5}^{Glu4} \bar{\upsilon}(p_2)\gamma(\hat{q}+\hat{p}_2)\hat{\varepsilon}^*(k_2)(\hat{k}_2-\hat{p}_2-\hat{q})\hat{\varepsilon}^*(k_1)(\hat{k}_1+\hat{k}_2-\hat{p}_2-\hat{q})\gamma u(p_1)}{144\pi^4 q^2(p_2+q)^2(q+p_2-k_2)^2(p_2+q-k_1-k_2)^2},$$

$$M_{11} = \frac{ee_q g_s^3 T_{Col5Col1}^{Glu5} T_{Col2Col5}^{Glu6} f^{Glu4Glu5Glu6}(2p_2\varepsilon^*(k_1)-k_1\varepsilon^*(k_1)-2k_2\varepsilon^*(k_1)+2q\varepsilon^*(k_1))\bar{\upsilon}(p_2)\gamma(\hat{q}-\hat{k}_2)\hat{\varepsilon}^*(k_2)\hat{q}\gamma u(p_1)}{144\pi^4 q^2(q-k_2)^2(p_2-k_2+q)^2(p_2+q-k_1-k_2)^2} +$$

$$+\frac{ee_q g_s^3 T_{Col5Col1}^{Glu5} T_{Col2Col5}^{Glu6} f^{Glu4Glu5Glu6}\bar{\upsilon}(p_2)(2\hat{k}_1+\hat{k}_2-\hat{p}_2-\hat{q})(\hat{q}-\hat{k}_2)\hat{\varepsilon}^*(k_2)\hat{q}\hat{\varepsilon}^*(k_1)u(p_1)}{144\pi^4 q^2(q-k_2)^2(p_2-k_2+q)^2(p_2+q-k_1-k_2)^2} +$$

$$+\frac{ee_q g_s^3 T_{Col5Col1}^{Glu5} T_{Col2Col5}^{Glu6} f^{Glu4Glu5Glu6}\bar{\upsilon}(p_2)\hat{\varepsilon}^*(k_1)(\hat{q}-\hat{k}_2)\hat{\varepsilon}^*(k_2)\hat{q}(\hat{k}_2-\hat{k}_1-\hat{p}_2-\hat{q})u(p_1)}{144\pi^4 q^2(q-k_2)^2(p_2-k_2+q)^2(p_2+q-k_1-k_2)^2},$$

$$M_{12} = \frac{iee_q g_s^3 T_{Col6Col1}^{Glu5} T_{Col2Col5}^{Glu4} T_{Col5Col6}^{Glu5} \bar{\upsilon}(p_2)\hat{\varepsilon}^*(k_1)(\hat{k}_1-\hat{p}_2)\gamma\hat{q}\gamma(\hat{k}_1-\hat{p}_2)\hat{\varepsilon}^*(k_2)u(p_1)}{144\pi^4 q^2(p_2-k_1)^4(p_2+q-k_1)^2},$$

$$M_{13} = \frac{iee_q g_s^3 T_{Col5Col1}^{Glu4} T_{Col2Col6}^{Glu5} T_{Col6Col5}^{Glu5} \bar{\upsilon}(p_2)\hat{\varepsilon}^*(k_2)(\hat{k}_2-\hat{p}_2)\gamma\hat{q}\gamma(\hat{k}_2-\hat{p}_2)\hat{\varepsilon}^*(k_1)u(p_1)}{144\pi^4 q^2(p_2-k_2)^4(p_2+q-k_2)^2}.$$